\documentclass[review,times]{elsarticle}
\usepackage[a4paper,left=1in,right=1in,top=1in,bottom=1in,footskip=.25in]{geometry}

\usepackage{moreverb}
\usepackage{graphicx,amsmath}
\usepackage{mathrsfs}
\usepackage{latexsym}
\usepackage{subfigure}
\usepackage{float} 
\usepackage{color}
\usepackage{array}

\usepackage{bm}
\usepackage{enumitem}
\usepackage{url}
\usepackage{lineno}
\usepackage{siunitx}

\biboptions{sort&compress}

\usepackage{amsthm}
\usepackage{multirow}
\usepackage[rgb]{xcolor}
\usepackage{pgfplots} 
\usepackage{seqsplit}

\usepackage{caption}
\makeatletter
\ams@newcommand{\vardot}[2]{%
  {\mathop{#2\kern0pt}\limits^{\vbox to-1.4\ex@{\kern-\tw@\ex@
   \hbox{\normalfont\multido{}{#1}{.}}\vss}}}}
\makeatother

\usepackage{tikz}
\usepackage{ifthen}
\usepackage{tikz-3dplot}
\usepackage{anyfontsize}
\usetikzlibrary{matrix}
\usetikzlibrary{calc}
\usetikzlibrary{graphs,graphs.standard,quotes,angles}
\usetikzlibrary{decorations,decorations.markings,decorations.text}
\usetikzlibrary{patterns}
\usetikzlibrary{shapes,arrows,fit,positioning,shadows}
\usetikzlibrary{decorations.pathreplacing}
\usetikzlibrary{pgfplots.groupplots}
\usepgfplotslibrary{polar}
\usetikzlibrary{mindmap}
\usetikzlibrary{arrows.meta, chains, shapes.multipart}
\usepgfplotslibrary{fillbetween}
\usepgfplotslibrary{statistics}

\usetikzlibrary{external}
\tikzset{external/system call={pdflatex \tikzexternalcheckshellescape 
                                        -halt-on-error
                                        -interaction=batchmode 
                                        -jobname "\image" "\texsource"
                                        && pdftops -eps "\image.pdf"}}
\pgfplotsset{every tick label/.append style={font=\tiny}}

\DeclareMathOperator*{\argminA}{arg\,min} 

\usepackage{cancel}
\usepackage{ulem}

\def \beq {\begin{equation}}
\def \eeq {\end{equation}}
\def \ba {\begin{array}}
\def \ea {\end{array}}
\def \dis {\displaystyle}
\def\RR{\mathbb{R}}
\def\NN{\mathbb{N}}
\def\EE{\mathbb{E}}
\def\cal{\mathcal}
\def\om{\omega}
\def\eps{\varepsilon}
\def\tb{\textbf}
\newcommand{\mbf}{\mathbf}
\DeclareMathOperator*{\argmaxA}{arg\,max}
\pgfplotsset{compat=newest} \pgfplotsset{plot coordinates/math parser=false}

\newproof{pf}{Proof}

\biboptions{numbers,sort&compress}

\journal{Applied Mathematical Modelling}

\usepackage{etoolbox}

\begin{document}

\begin{frontmatter}


\title{Multielement polynomial chaos Kriging-based
metamodelling for Bayesian inference of non-smooth systems}

\author[label1]{J.C. Garc\'{i}a-Merino\corref{cor1}}
\ead{jcgarcia@unex.es}

\author[label1]{C. Calvo-Jurado}

\author[label3]{E. Mart\'{i}nez-Pa\~neda}

\author[label3,label2]{E. Garc\'{i}a-Mac\'{i}as}

\address[label1]{Department of Mathematics, School of Technology, 10003, C\'aceres (Spain)}
\address[label3]{Department of Civil and Environmental Engineering. Imperial College London SW7 2AZ, (UK)}
\address[label2]{Department of Structural Mechanics and Hydraulic Engineering, Campus Universitario de Fuentenueva (Edificio Polit\'ecnico) 18071 Granada (Spain)}

\cortext[cor1]{Corresponding author. Department of Mathematics, School of Technology, 10003, C\'aceres (Spain)}

\date{}

\begin{abstract}
This paper presents a surrogate modelling technique based on domain partitioning for Bayesian parameter inference of highly nonlinear engineering models. In order to alleviate the computational burden typically involved in Bayesian inference applications, a multielement Polynomial Chaos Expansion based Kriging  metamodel is proposed. The developed surrogate model combines in a piecewise function an array of local Polynomial Chaos based Kriging metamodels constructed on a finite set of non-overlapping subdomains of the stochastic input space. Therewith, the presence of non-smoothness in the response of the forward model (e.g.~ nonlinearities and sparseness) can be reproduced by the proposed metamodel with minimum computational costs owing to its local adaptation capabilities. The model parameter inference is conducted through a Markov chain Monte Carlo approach comprising adaptive exploration and delayed rejection. The efficiency and accuracy of the proposed approach are validated through two case studies, including an analytical benchmark and a numerical case study. The latter relates the partial differential equation governing the hydrogen diffusion phenomenon of metallic materials in Thermal Desorption Spectroscopy tests.
\end{abstract}



\end{frontmatter}



\section{Introduction}

The widespread use of high performance computing (HPC) technologies has enabled the increasingly frequent adoption of computationally intensive numerical models in a myriad of disciplines in academia and industry. Such high fidelity models allow conducting virtual testing of engineering systems, avoiding the technical limitations and minimizing the costs associated with traditional experimental testing. Nevertheless, formidable challenges still arise when implementing these models into computationally demanding studies such as optimization~\cite{Chen2021}, sensitivity analysis~\cite{sun2020}, model identification and calibration~\cite{GarciaMacias2020}, reliability analysis~\cite{Hong2021}, or robust design~\cite{Yuan2021}. As a solution, a variety of surrogate models or metamodels have been proposed in the literature in recent years. Nevertheless, despite considerable the advances in the field, there remain open research challenges for their extensive use such as handling highly nonlinear model responses, limited training datasets and, in general, the online implementation of surrogate models to enable decision-making in engineering systems~\cite{Stork2020,Hao2021}. In particular, the development of real time surrogate model-based parameter estimation approaches draws high interest in frontier research fields such as digital twins development~\cite{Chakraborty2021} and smart maintenance of engineering systems~\cite{Sun2021}.
 
In their broadest sense, surrogate models are computationally light  black box representations of resource-intensive models. Surrogate modelling methods can be generally classified into three categories~\cite{Asher2015}: (i) projection-based methods or reduced-order models (ROMs)~\cite{Gooijer2021}; (ii) multi-fidelity methods~\cite{Zhang2022}; and (iii) data-driven or responsive surface methods (RSMs). Projection-based techniques project the governing equations of the original model onto a low-dimensional subspace. Hence, although ROMs have the advantage of retaining the physics underlying the model, these are limited to situations where access to the governing equations is granted, which is not the case in many practical applications when using commercial software. Multi-fidelity methods are built by simplifying the underlying physics or reducing the numerical resolution. These methods may also render some difficulties, being highly case-dependent and requiring specialized expertise to find a suitable trade-off between prediction accuracy and computational burden. Such difficulties have fostered rapid developments of RSMs in recent years as a non-intrusive technique with great flexibility in a wide range of applications. The key advantage of these methods relies in the fact that the forward model does not need to be modified and, therefore, it can be essentially treated as a black box~\cite{shi2019}. Among the broad variety of RSMs available in the literature, some of the most popular ones are Kriging~\cite{kleijnen2009krigingbasic}, radial basis functions (RBF)~\cite{buhmann2000radial}, support vector regression (SVR)~\cite{smola2004vector}, artificial neural networks (ANN)~\cite{Kingston2011}, Gaussian process (GP) regression~\cite{Schulz2018}, polynomial chaos expansions (PCE)~\cite{sun2020}  and Polynomial Chaos Expansion based Kriging (PCK)~\cite{Sch-2015}. These models are trained by exploiting a set of realizations of  parameters of interest of the model, called the experimental design (ED), and the corresponding model evaluations, also called quantities of interest (QoI). Therefore, the accuracy of data-driven surrogate models is highly determined by the dimensions of the design space and the number and distribution of the training samples in the ED~\cite{Alizadeh2020}. In engineering practice, obtaining the ED constitutes the most time-consuming part since it requires the evaluation of the computationally intensive forward model at each sample point. Choosing high quality EDs is thus critical to achieve high accuracy in the surrogate model construction with the least possible number of training samples. Sampling approaches can generally be divided into static (one-shot) and sequential methods. One-shot sampling generates the training sample points in one single step, and common approaches include fractional designs and orthogonal arrays~\cite{Queipo2005}. While these techniques offer easy implementation and minimal computational cost, the determination of the optimal sample size may be troublesome when the behaviour of the forward model is unknown. To minimize such difficulties, a number of sequential sampling strategies have been introduced, including adaptive and space-filling sequential methods~\cite{Fuh-2021}. On one hand, space-filling sequential designs such as Latin Hypercube Sampling (LHS), Sobol, Hammersley and Halton sampling~\cite{Garud2017} generate samples iteratively to attain good coverage of the parametric domain. On the other hand, adaptive sampling techniques iteratively draw new samples in regions of the parametric space with large prediction errors, enabling to account for local refinements in the ED (refer to references~\cite{Liu2018,Fuh-2021} for a thorough state-of-the-art review).

A second major challenge of non-intrusive surrogate models regards the difficulties involved in the fitting of non-smooth  models exhibiting unsteadiness, sparseness or large perturbations. In these cases, adaptive solutions accommodating local relevant refinements of the response surface are required. A large volume of research has been conducted in the last decade to address this issue, giving origin to a number of advanced surrogate modelling techniques such as multi-resolution generalized PCE~\cite{Mai-2004}, domain partitioning~\cite{menafoglio2018}, sparse grid collocation~\cite{Resmini2016}, Voronoi tesselations~\cite{Mattis2019}, local search at trust regions~\cite{Ong2003}, clustering-based partitioning~\cite{Liem2015}, ensembles of surrogate models~\cite{Teixeira2021}, multi-element generalized polynomial chaos (ME-gPC) method \cite{Resmini2016} and multi-element probabilistic collocation methods (ME-PCM)~\cite{Foo2008}. In this light,  different partitioning techniques  can be found in the literature in  the context of the pure GPs~\cite{Ras-2001, Kon-2019}, PCE~\cite{Mar-2021}, and PCE-based mapping of likelihood functions in parameter inference applications~\cite{Wag-2021,Mai-2004}. In those works, different domain partition criteria were proposed based on dissimilarity measurements between regions~\cite{Kon-2019}, data density~\cite{Ras-2001}, or maximum residual differences~\cite{Mar-2021,Wag-2021}.  In general, these approaches usually generate series of subdomains  where the non-smooth response surface can be assumed locally smooth, thus enabling the definition of local surrogate models contributing to the global response in a piecewise fashion.

The development of cost-efficient surrogate models opens vast new opportunities for real-time parameter estimation applications. Probabilistic Bayesian approaches are particularly attractive owing to their ability to assess the effects of uncertainties on the model parameters and the derived response predictions, as well as their robustness to noise pollution and efficiency to deal with ill-conditioning and ill-posedness. Such excellent features have fostered their implementation in multiple fields such as structural identification~\cite{Huang2019}, geotechnical problems~\cite{Liu2019}, material characterization~\cite{Emery2016}, and bioengineering~\cite{Hauseux2018}, just to mention a few. In general, Bayesian parameter estimation approaches exploit experimental data to infer the posterior probability distribution functions (PDFs) of certain unknown model parameters through the Bayes' theorem. Nonetheless, the direct evaluation of the posterior PDFs requires solving the possibly high-dimensional integral related to the evidence of the model. Therefore, except for some trivial cases, posterior PDFs often need to be approximated numerically. Markov chain methods constitute the most widespread set of techniques to extract series of samples to estimate the posterior PDFs, allowing to sample from a large class of high-dimensional distributions. Popular procedures for Markov chain Monte Carlo (MCMC) sampling are the Metropolis-Hastings~\cite{Dwi-2018}
and Gibbs algorithms~\cite{Cheung2017}. The basic idea of these techniques is to construct a Markov chain with a stationary distribution resembling the posterior distribution, in such a way that a sample of the joint PDF of the model parameters can be obtained by collecting the states of the chain. These methods guarantee asymptotic convergence to the exact PDF, although a considerably large number of iterations are typically required to achieve convergence, which compromises the computational efficiency of the inference~\cite{che2021}. To alleviate the computational burden in classical MCMC techniques, a variety of more efficient sampling algorithms have been proposed in recent years, including Sequential Monte Carlo (SMC)~\cite{DelMoral2006}, Transitional MCMC~\cite{Ching2007}, and Bayesian broad learning~\cite{Yin2020}. Notwithstanding these advances, their elevated computational cost remains a critical limitation when high-fidelity models are considered in the inference problem. Herein is where surrogate models offer an efficient solution to conduct cost-efficient Bayesian inference while retaining the accuracy of high-fidelity models. The enormous potentials of this approach are evidenced by the increasing number of research studies reported in recent years. It is worth noting the work by Schneider \textsl{et al}.~\cite{Schneider2022} who proposed a Bayesian procedure using rational PCE metamodels of the response of dynamic systems in the frequency domain, and demonstrated its effectiveness for the identification of a cross-laminated timber plate. Xing \textsl{et al}.~\cite{Xing2021} developed an additive GP model for multi-fidelity surrogate modelling inserted into a non-parametric Bayesian approach with a closed-form solution for the predictive posterior PDF. The accuracy and flexibility of their approach were validated with several benchmark problems, including the identification of a solid oxide fuel cell model and an elbow-shape pipe under turbulent mixing flow conditions. Ierimonti and co-authors~\cite{Ierimonti2021} proposed a Kriging-based conjugate Bayesian identification methodology for online damage identification of an instrumented monumental building, the Consoli Palace in Gubbio (Italy). Del Val \textsl{et al}.~\cite{del2022} implemented an MCMC approach to infer the catalytic recombination parameters of reusable thermal protection materials exploiting plasma wind tunnel measurements. Interestingly, instead of bypassing a high-fidelity model, those authors approximated the likelihood function in the Bayesian inference using a GP surrogate model to accelerate the parameters estimation. 


In light of the literature review above, it is apparent that the fields of surrogate modelling and its applications for fast parameter estimation have experienced considerable advances in recent years. Nonetheless, light metamodels capable of bypassing highly nonlinear models for online parameter estimation are yet to be fully developed. In this context, the present work proposes a novel multielement surrogate model extending the PCK proposed in \cite{Sch-2015} for online Bayesian parameter inference of non-smooth models. With the aim of accommodating nonlinear behaviours in the forward model while retaining flexibility and minimal computational cost, a simple regular block partitioning approach is implemented. On this basis, the space domain is partitioned into distinct subregions with splitting  directions chosen on the basis of a preliminary sensitivity analysis, prioritising the division of those variables with the highest sensitivities. In the present study, Sobol's indices over the full domain have been considered for this purpose, since they can be readily obtained as a by-product of PCE \cite{Bla-2011}. With respect to the construction of the local PCK surrogate models the optimal order of the polynomials in the PCE is automatically identified by a model selection technique for sparse linear models, the least-angle regression (LAR) algorithm set out by Efron \textsl{et al}.~\cite{Efr-2004}. Then, the optimal PCE is inserted into a Kriging predictor as the trend term, while the stochastic term is fitted through a genetic algorithm (GA) global optimization approach. In this regard, the domain space is split into a discrete number of subsets where local surrogate models are constructed. Then, the global model response is obtained by combining the local surrogate models in a piecewise fashion. In order to minimize the computational burden in the construction of the surrogate models, the ED is obtained with the adaptive Monte Carlo-Intersite-proj-th (MIPT) approach~\cite{Fuh-2021}. Finally, the surrogate model is used for Bayesian parameter estimation using a cost-efficient adaptive MCMC with delayed rejection (DRAM) algorithm developed by Haario \textsl{et al}.~\cite{Haa-2006}. The effectiveness of the proposed approach is validated through two benchmark case studies: (i) an analytical benchmark; (ii) and a  partial differential equation (PDE) for thermal desorption spectroscopy (TDS) experiments of hydrogen in metals. The latter represents a formidable example of a model exhibiting a non-smooth behaviour in the shape of nonlinearities and unsteadiness. The presented results demonstrate the effectiveness of the proposed approach for conducting online Bayesian parameters identification, proving robustness to the presence of highly nonlinear behaviours and multimodality in the posterior distributions.

The remainder of this paper is organized as follows. Section~\ref{Sect2} outlines the theoretical formulation of the proposed approach. Section~\ref{Sect3} presents the numerical results and discussion. In particular, two case studies are investigated, namely a benchmark analytical model and a numerical model of the TDS analysis of hydrogen desorption in metals. Finally, Section~\ref{Sectconc} discusses the contributions of this work and presents the main concluding remarks.


\section{Theoretical formulation}\label{Sect2}

\begin{figure}[H]
    \centering
        \includegraphics[scale=0.9]{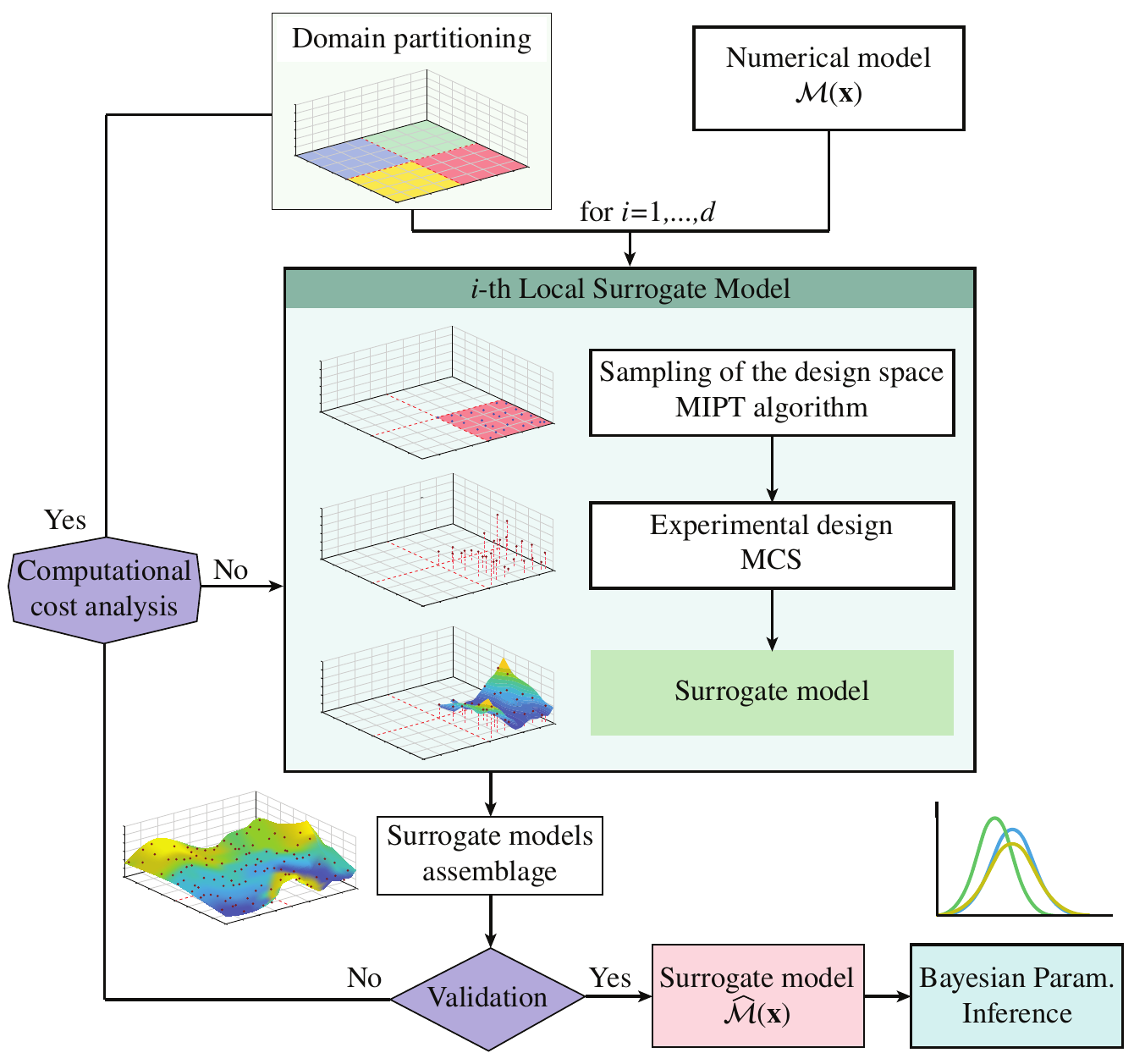}
    \caption{Flowchart of the proposed PCK metamodel to perform Bayesian Inference.}
    \label{flowchart}
\end{figure}

The main purpose of this section is to present the theoretical fundamentals of the proposed surrogate model-based Bayesian inference approach. The general methodology is sketched in Fig.~\ref{flowchart} and comprises four main steps, namely (i) domain partitioning; (ii) construction of the local surrogate models; (iii) surrogate models assemblage; and (iv) validation.

\subsection{Surrogate modelling: Polynomial Chaos Expansion based Kriging}

Let $(\Omega,\Sigma,\mu)$ be a probability space, with $\Omega\subset \RR^M$ denoting the event space equipped with a $\sigma$-algebra $\Sigma$ of subsets of $\Omega$ and a probability measure $\mu$ such that $\mu\left( \Omega \right)=1$. Let ${\cal M}:\Omega\subset \RR^M\to\RR$ be a computational model mapping a vector of input variables $\textbf{x} = \left[x_1,\ldots,x_M\right]^{\textrm{T}}$ into the output variable or quantity of interest $y\in\RR$. The goal of surrogate modelling is to approximate the forward model ${\mathcal{M}}$, which is typically computationally intensive, by a computationally inexpensive function $\hat{\mathcal{M}}$. In this paper,  PCK metamodels combining PCE and Kriging are adopted. In this light, PCE is used to approximate the global behaviour of the computational model ${\cal M}$ while the Kriging metamodel captures its local behaviour.


\subsubsection{Polynomial Chaos Expansion}\label{PCESect}

Assume that the input vector ${\tb x}\in \Omega$ is constituted by $M$ independent random variables components $\{x_i\}_{i=1}^M$ with probability density functions $\mu_{X_i}$. Then, PCE represents the output response $y\in \RR$ as the infinite expansion of $\mathcal{M}$ over an orthonormal basis of multivariate polynomials $\Psi_{\bm\alpha}$ as:

\beq\label{expansion}
y=\mathcal{M}\left(\textbf{x}\right)=\sum_{\bm{\alpha} \in \mathbb{N}^M} a_{\bm{\alpha}} \Psi_{\bm{\alpha}}(\textbf{x}),
\eeq

\noindent where $a_{\bm{\alpha}}$, ${\bm{\alpha}}=(\alpha_1,\ldots,\alpha_M)$, $\alpha_i\in\NN$ are the coefficients of the expansion. Orthonormal basis families associated with a variety of standard distributions can be found in reference
~\cite{Pho-2015}. Multivariate polynomials $\Psi_{\bm{\alpha}}$ can be expressed in terms of a family of univariate polynomials $\left\{\psi_j^{(i)}, \, j \in \mathbb{N}\right\}$ as $\Psi_{\bm{\alpha}}(\textbf{x}) = \prod_{i=1}^M \psi_{\alpha_i}^{(i)}(x_i)$. Polynomials $\psi_j^{(i)}$ are also orthonormal with respect to the marginal distribution, that is:

\beq\label{expectt}
\mathbb{E} \left[ \psi_j^{(i)}(x_i), \, \psi_k^{(i)}(x_i) \right]= \int \psi_j^{(i)}(u) \, \psi_k^{(i)}(u) \, \mu_{x_i}(u) \, \textrm{d}u=\delta_{j,k},
\eeq

\noindent with $\delta_{j,k}$ being the Dirac Delta function. Random variable ${\tb x}$ induces the probability measure $\mu_{\tb x}$ on $\left(\RR^M,{\cal B}\left(\RR^M\right)\right)$, with ${\cal B}\left(\RR\right)$ denoting the Borel $\sigma$-algebra $\mu_{\tb x}(B)=\mu\{\om\in\Omega: \, \tb{x}(\om)\in B\}, \, B\in{\cal B}$. The PDF of $\tb{x}$ is given by the probability measure $\mu_{\tb x}$ as $\mu_{\tb x} (\tb{u}) = \mu ({\tb x} \leq \tb{u}), \;  \tb{u} \in \RR^M$. Note that, since the components of ${\tb{x}}$ are independent, the joint probability function $\mu_{\tb x} (\tb{u})$ is given by the product of the marginal PDFs of $x_i$, i.e.~$\mu_{\textbf{x}}\left(\textbf{u}\right) = \prod_{i=1}^M \mu_{x_i}\left(u_i\right),$ $\tb{u}=(u_1,\ldots,u_M)\in\RR^M$. Therefore, Eq.~(\ref{expectt}) can be written in a more compact form as:

\begin{equation}
\mathbb{E} \left[ \Psi_{\bm \alpha}({\tb{x}}), \, \Psi_{\bm \beta}({\tb{x}}) \right]= \int \Psi_{\bm \alpha}({\tb{u}}) \Psi_{\bm \alpha}({\tb{u}}) \, \mu_{\tb{x}}(\tb{u})\ \textrm{d}\tb{u} = \bm{\delta}_{\bm{\alpha}, \bm{\beta}},\quad{\bm\alpha},\,{\bm\beta}\in\NN^M.
\end{equation}

Although the expression in Eq.~(\ref{expansion}) is exact for an infinite number of terms, in practice only a finite number can be computed and a certain truncation scheme needs to be adopted. One of the simplest approach consists in selecting all the polynomials whose total degree $|\bm\alpha|=\dis\sum_{i=1}^M \alpha_i$ belongs to the set:

\begin{equation}
\mathcal{A}^{M,p}=\left\{\bm{\alpha} \in  \mathbb{N}^M: 0 \leq \left| \bm{\alpha}\right|\leq p \right\},\quad \mbox{with} \quad \textrm{card} \,\mathcal{A}^{M,p} = \left(\begin{matrix}
M+p\\
p\\
\end{matrix}
\right)=\frac{\left(M+p\right)!}{M!p!}.
\end{equation}

For high-dimensional and non-linear problems, this truncation procedure usually leads to large numbers of polynomial coefficients and considerable computational burdens. Nevertheless, it is often observed in many practical applications that coefficients corresponding to high interaction terms between the input variables are close to zero, a phenomenon that is also known as the {\it sparsity-of-effect principle}~\cite{Bla-2011}. To alleviate this, an hyperbolic truncation scheme can be adopted. This approach selects all multi-indices with $q$-norm $\|\bm{\alpha}\|_q=\left(\dis\sum_{i=1}^M\alpha_i^q \right)^\frac{1}{q}$ less than or equal to a certain model order $p$, i.e.~$\mathcal{A}^{M,p,q} = \left\{ \bm{\alpha} \in \mathbb{N}^M: \, \left\| \bm{\alpha} \right\|_q \leq p\right\}$. Note that the expansion tends to only maintain univariate polynomials as the $q$-norm decreases, thus achieving a reduction in the computational cost of the metamodel. Nonetheless, the number of terms may remain elevated since the potential sparseness in the coefficients is not being actually assessed. Following the work by Blatman and Sudret~\cite{Bla-2011}, the LAR algorithm~\cite{Efr-2004} is adopted to further reduce the number of polynomial coefficients. In the context of PCE, LAR constructs a set of expansions incorporating an increasing number of basis polynomials $\Psi_{\bm{\alpha}}$, from 1 to ${\mathcal P}=\textrm{card}\left(\mathcal{A}^{M,p,q}\right)$. The resulting sequence of index sets is used to construct a family of expansions with decreasing sparseness. Finally, a cross-validation procedure can be implemented to select the best metamodel among the obtained family of expansions. In particular, the Bayesian Information Criterion (BIC) is adopted in this work. Once the optimal index set is selected, the expansion coefficients ${\bm a}=\{{\bm a_\alpha}, \, {\bm \alpha}\in \mathcal{A}^{M,p}\subset\NN^M\}$ are obtained by minimizing the expectation of the least squared error:

\beq\label{leastsq} 
{\bm a}= {\rm arg\dis\min_{\mbox{$\bm a$} \, \in \, \RR^{P}}}\EE \left[\left({\cal M}({\tb{x}})-\dis\sum_{{\bm \alpha}\,\in\,{\cal A}^{M,p,q}}{a}_{\bm \alpha}{\Psi}_{\bm\alpha}(\tb{x})\right)^2\right].
\eeq

In practice, Eq.~(\ref{leastsq}) is calibrated on a set of $N$ realizations $\Xi=\left\{\tb{x}^{(1)},\ldots,\tb{x}^{(N)}\right\}$ of the input variable $\tb{x}$ forming the ED. In order to obtain a representative ED, the adaptive MIPT sampling method~\cite{Fuh-2021} is adopted in this work. Then, the expectation operator in Eq.~(\ref{leastsq}) is replaced by its discretized version using the ED realizations as follows:

\beq\label{leastsqdis} 
{\bm a}= {\rm arg\dis\min_{\mbox{\bm $a$} \, \in \, \RR^P}}\dis\frac{1}{N}\dis\sum_{i=1}^N\left({\cal M}\left(\tb{x}^{(i)}\right)-\dis\sum_{{\bm\alpha} \, \in \, {\cal A}^{M,p,q}}{\bm a}_{\bm \alpha}\Psi_{\bm\alpha}\left(\tb{x}^{(i)}\right)\right)^2.
\eeq

Denoting the realizations of the output variable $y$ by ${\tb y}=\{y^{(1)}={\cal M}(\tb{x}^{(1)}),\ldots, y^{(N)}={\cal M}(\tb{x}^{(N)}) \}^\textrm{T}$, the solution of the optimization problem in Eq.~(\ref{leastsqdis}) reads:

\begin{equation}\label{sol}
\hat{\bm{a}}=\left(\bm{\Theta}^{\textrm{T}} \bm{\Theta} \right)^{-1}\bm{\Theta}^{\textrm{T}}{\bm y}, \qquad \bm{\Theta}=(\bm{\Theta}_{ij})=\left[\psi_{j}(\bm{x}^{(i)}) \right]_{i=1,\ldots,N}^{j=1,\ldots,P},
\end{equation}

\noindent where $\bm{\Theta}$ denotes the information matrix calculated from the evaluation of the basis polynomials on $\Xi$. For the least-square minimization problem in Eq.~(\ref{leastsqdis}) to be well posed, the size of the ED is usually selected according to the heuristic rule $N\approx 2\cdot P$ or $3\cdot P$~\cite{Bla-2011}. Once Eq.~(\ref{leastsqdis}) has been solved, the predictions of the PCE surrogate model can be obtained as:

\beq\label{PCE} 
\hat{y}=\hat{\mathcal{M}}_{PCE}(\tb{x}) = \sum_{\bm{\alpha} \in \mathcal{A}} \hat{\bm{a}}_{\bm{\alpha}}\Psi_{\bm{\alpha}}(\tb{x}).
\eeq


\subsubsection{Polynomial Chaos Expansion based Kriging (PCK)}\label{PCK}

The Kriging method assumes that the response of a computational model ${\cal M}({\tb{x}})$ is modelled by the sum of a stochastic random process $\cal{Z}(\textbf{x})$ and a regression model $\mathcal{T}(\textbf{x})$, also called trend, in the form~\cite{Fuh-2021}:


\begin{equation}\label{Krig1}
\hat{\mathcal{M}}(\textbf{x})=\mathcal{T}(\textbf{x})+\mathcal{Z}(\textbf{x}).
\end{equation}

The stochastic component in Eq.~(\ref{Krig1}) is fully determined by the covariance function~\cite{Sac-1989}:

\begin{equation}\label{Krig2}
\textrm{Cov}\left(\mathcal{Z}\left(\textbf{x}\right),\mathcal{Z}\left(\textbf{x}'\right)\right) = \mathbb{E} \left[\mathcal{Z}(\textbf{x})\mathcal{Z}(\textbf{x}') \right] = \sigma^2 \, R\left(\left| \textbf{x}-\textbf{x}' \right|;\bm{\theta}\right),
\end{equation}

\noindent with $\sigma^2$ being the process variance, and $R\left(\left| \textbf{x}-\textbf{x}' \right|;\bm{\theta}\right)$ an auto-correlation function \cite{Ras-2006} between two input sample points $\textbf{x}$ and $\textbf{x}$' that depends on certain hyper-parameters $\bm{\theta}$ to be computed. In this work, the Gaussian correlation function is adopted as:

\begin{equation}\label{R}
R\left(\textbf{x},\textbf{x}',\bm{\theta} \right) = \prod_{\ell=1}^M \exp \left[ -\theta_\ell \left(x_\ell-{x'}_\ell \right)^2 \right] \, .
\end{equation}

The trend term of the Kriging model in Eq.~(\ref{Krig1}) interpolates the forward model evaluations at the ED, while the local variability is captured by the stochastic process. Depending on the form of the trend, three different versions of Kriging are typically referred to in the literature~\cite{Fuh-2021}, including simple, ordinary and universal Kriging, which respectively correspond to polynomials of degrees $0$, $1$ and $N$. In this work, with the aim of combining the excellent global approximation capabilities of the PCE previously introduced in Section~\ref{PCESect}, the sparse PC expansion obtained by LAR is introduced in the shape of the trend term in Eq.~(\ref{Krig1}). The resulting PCK metamodel reads:

\begin{equation}\label{Krig3}
\hat{\mathcal{M}}_{PCK}(\textbf{x})=\hat{\cal M}_{PCE}(\tb{x})+\mathcal{Z}(\textbf{x})=\sum_{\bm{\alpha} \, \in \, \mathcal{A}} \hat{\bm{a}}_{\bm{\alpha}} {\color{red}\Psi}_{\bm{\alpha}}(\textbf{x})+\mathcal{Z}(\textbf{x}).
\end{equation}

The construction of the PCK metamodel in Eq.~(\ref{Krig3}) consists in two steps. Firstly, the optimal set of orthonormal polynomials ${\bm\Psi}_{\bm{\alpha}}$ (for $\bm{\alpha}\in {\cal A}$ the truncation set) is obtained by LAR as indicated in Section~\ref{PCESect}. Secondly, the calculation of hyperparameters $\hat{\bm{\theta}}$ and the polynomial coefficients and the process variance $\{\bm{a}(\hat{\bm{\theta}}),\,\sigma^2(\hat{\bm{\theta}})\}$ are obtained. The optimal correlation parameters $\hat{\bm{\theta}}$ can be determined by the Maximum-Likelihood-Estimation (ML) through the following minimization problem~\cite{Cha-2017}:

\begin{equation}\label{ML}
\hat{\bm{\theta}} = \argminA_{\bm{\theta}} \left[\frac{1}{N} \left( \tb{y} -\bm{\Theta}\bm{a} \right)^{\textrm{T}} \textbf{R}^{-1} \left( \tb{y} -\bm{\Theta}\bm{a} \right) \left( \textrm{det} \, \textbf{R} \right)^{1/N} \right].
\end{equation}

In order to solve the optimization problem in Eq.~(\ref{ML}), local optimization algorithms such as gradient-based methods are often used. Nonetheless, a major drawback of these techniques relates the troublesome identification of global maxima/minima, being possible to get stuck in local maxima/minima. To avoid this, a global genetic algorithm optimization procedure is used in this work. Since the correlation matrix is symmetric and positive definite, its inverse in Eq.~(\ref{ML}) is computed by Cholesky decomposition. Then, once $\hat{\bm{\theta}}$ is computed, the polynomial coefficients and the process variance $\{\bm{a}(\hat{\bm{\theta}}),\, \sigma^2(\hat{\bm{\theta}})\}$ are calculated using the Empirical Best Linear Unbiased Estimator (BLUE) as~\cite{Kle-1992}:

\begin{equation}\label{krparm1}
\bm{a}\left(\hat{\bm{\theta}}\right) = \left( \bm{\Theta}^{\textrm{T}} \textbf{R}^{-1} \bm{\Theta} \right)^{-1}\bm{\Theta}^\textrm{T}\textbf{R}^{-1}\tb{y}, \quad \sigma^2\left(\hat{\bm{\theta}}\right) = \frac{1}{N} \left(\tb{y} -\bm{\Theta}\bm{a}\right)^{\textrm{T}} \textbf{R}^{-1} \left( \tb{y} -\bm{\Theta}\bm{a} \right),
\end{equation}

\noindent where ${R}_{ij}={ R}\left(\left| \tb{x}^{(i)}-\tb{x}^{(j)} \right|; \hat{\bm{\theta}} \right)$ is the correlation matrix and $\bm{\Theta}_{ij}=\psi_{j}\left(\tb{x}^{(i)} \right)$ the information matrix evaluated at all the samples of the ED.

\paragraph{Effective explorative sampling}

With the aim of generating representative EDs, the MIPT algorithm is adopted as a computationally efficient and easily implementable adaptive sampling technique. The main advantage of this technique compared to space-filling techniques such as LHS regards its ability to avoid local clustering of points which may consequently lead to numerical instabilities in the inverse of the Kriging correlation matrix in Eq.~(\ref{krparm1}). This exploration distance-based sampling method iteratively augments the ED by adding new sampling points with maximum distance with respect to the data population in the ED among a large set of $N_c$ random Monte-Carlo candidates. Specifically, among the candidates set $\cal {C}=\{\bm{\xi}^{(1)},\,\bm{\xi}^{(2)},\ldots,\bm{\xi}^{(N_c)}\}$, a new sample $\tb{x}^{(N+1)}$ is chosen by solving the following optimization problem:

\beq\label{adaptive}
\tb{x}^{(N+1)}=\argmaxA_{\bm{\xi}^\ast \, \in \, \cal {C}} \left({\displaystyle\min}_{\tb{x}^{(i)}, \, {i=1,\ldots,N}} \left\| \bm{\xi}^\ast-\tb{x}^{(i)} \right\|_2\right),
\eeq

\noindent with $\left\| \cdot \right\|_2$ the euclidean norm, i.e.~$\left\| \tb{x} \right\|_2=\left(\sum_{i=1}^M x_i^2\right)^{1/2}$.

\subsubsection{Multi-element surrogate model approach}\label{multi}

The previously presented PCK metamodel suffers from low convergence rates when the forward model $\mathcal{M}$ exhibits non-smoothness \cite{Pel-2014}. Thus, considerably large ED sizes are often required to achieve accurate predictions.  
This aspect undermines the computational efficiency of the PCE-based Kriging model, which is dominated by  the   $\mathcal{O}(N^3)$ complexity of the Kriging predictor. 
In turn, this implies long construction times or even memory overflow issues when  solving the optimization problem in Eq. (\ref{ML}). Moreover, the larger the size of the ED, the slower the evaluation of the corresponding metamodel, which reduces or vanishes the advantages of the surrogate approach. To address this issue, a multi-element PCK model inspired by the ME-gPC method by Wan and Karniadakis \cite{Wan-2006} is proposed in this work. This approach consists in the partitioning of the random input space into a finite set of non-overlapping subdomains,  the construction of a local PCK surrogate model in each one following the formulation in Section \ref{PCK} and, finally, assembling them into a piecewise function to obtain a global metamodel, as sketched in Fig. \ref{flowchart}.

In order to address the direction of the partitions, an approach relying on sensitivity analysis based on the Sobol's indices has been adopted in this work. Note that the Sobol's indices can be readily computed as a by-product of the PCE \cite{Bla-2011}. In this way, priority in the partitioning is given to the   direction of those parameters with highest sensitivity, i.e., those with the greatest effect on the variability of the quantity of interest $y$. On the other hand, in the following analyses the number of divisions have been a priori determined in order to compare models built with the same amount of information.

On this basis, for the random variable $\mathbf{x}: \Omega \rightarrow \mathcal{D}_{\mathbf{x}} \subset \mathbb{R}^{M}$,  a decomposition is defined as

\begin{equation}\label{decomp} {\mathcal  D}_{\textbf{x}}=\bigcup_{j\in{\mathcal  J}}{\mathcal
D}_{j},\quad {\mathcal  D}_{j}\cap{\mathcal 
D}_{j'}=\emptyset,\quad\mbox{if}\, j\neq j',  \end{equation}

\noindent where $\chi_{{\mathcal  D}_j}:\Omega \to\RR$ denotes the  indicator random variable:  $$ \chi_{{\mathcal  D}_j}(\tb x)=\left\{\begin{array}{ll}
1 & \mbox{ if } \; \tb x \in{\mathcal  D}_j\\ 
0 & \mbox{otherwise}
\end{array}\right..$$

In this way, the global model is defined in a piecewise fashion as:  
\begin{equation}\label{piece}
\hat{\mathcal M}_{PCK}({\textbf{x}})=\dis\sum_{j \, \in \, \mathcal J}\chi_{{\mathcal  D}_{j}}({\textbf{x}})\hat{\mathcal  M}^{j} ({\textbf{x}}).\end{equation}

As aforementioned, the number of partitions in this work is defined after a parametric analysis. Nevertheless, the previous formulation may be readily automated as follows. The splitting criterion of the domain is determined by a certain user-defined accuracy goal and a minimum number of samples ${\mathcal N}$ per region. Afterwards, the splitting process is performed iteratively from a PCK model built over the full parameter space $\Omega$. In case the target accuracy has not been reached, the space is split into two regions and the ED is enriched in each of these subdomains by the MIPT algorithm until there are ${\mathcal N}$ samples in each one. Note that, given the sequential nature of MIPT, the information of the previously extracted samples is not lost. If the accuracy goal is not reached yet, a new division of the space and a new enrichment of the ED are performed


\subsubsection{Surrogate model accuracy. Complexity analysis of the algorithm}
\label{MetricsAcc}

To evaluate accuracy of the developed metamodel, both local and global error metrics are considered. These metrics are computed by considering a validation set (VS) $\bm{\Lambda}=\{\bm{\xi}^{(1)},\ldots,\bm{\xi}^{(K)}\}$, $K\in \NN$, of the parameters space (independent of the ED). Denote by ${\bm{\Upsilon}}=\{{\upsilon}^{(1)}={\cal M}(\bm{\xi}^{(1)}), \ldots, {\upsilon}^{(K)}={\cal M}(\bm{\xi}^{(K)})\}$ and $\hat{\bm{\Upsilon}}=\{\hat{{\upsilon}}^{(1)}=\hat{{\cal M}}_{PCK}(\bm{\xi}^{(1)}), \ldots, \hat{{\upsilon}}^{(K)}=\hat{{\cal M}}_{PCK}(\bm{\xi}^{(K)})\}$ the outputs of the VS estimated by the forward model and the metamodel, respectively. Then, the accuracy of the surrogate model can be assessed by using the error metrics like those collected in Table~\ref{errormetrics}. In this table, $\bar{\bm{\Upsilon}}$ and $\mbf{\sigma}_{\bm{\Upsilon}}=\sqrt{\left(\sum_{i=1}^K\left(\bar{\bm{\Upsilon}}-{\upsilon}^{(i)}\right)^2\right)/\left(K-1\right)}$ denote the arithmetic mean and the quasi standard deviation of ${\bm{\Upsilon}}$, respectively. Term $\mbf{\sigma}_{ \bm\Upsilon\hat{\bm\Upsilon} }$ represents the covariance of $({\bm{\Upsilon}},\hat{\bm{\Upsilon}})$, and $\sigma^2_{\bm{\Upsilon}}$ and $\sigma^2_{\hat{\bm{\Upsilon}}}$ indicate the variance of  ${\bm{\Upsilon}}$ and $\hat{\bm{\Upsilon}}$, respectively. Note that the error metric NMAE in Table~\ref{errormetrics} provides a local estimation of accuracy, while NRMSE, NAAE, and $R^2$ represent global accuracy measures.

\begin{table}[H]
\newcommand\Tstrut{\rule{0pt}{0.6cm}}         
\newcommand\Bstrut{\rule[-0.4cm]{0pt}{0pt}}   
\newcommand\Tstrutb{\rule{0pt}{0.3cm}}         
\newcommand\Bstrutb{\rule[-0.15cm]{0pt}{0pt}}   
\footnotesize
\caption{Error metrics for the accuracy assessment of surrogate models over a validation set (VS) of size $K$ (Ref.~\cite{Mou-2018}).}\label{error_metrics}
\vspace{0.1cm}
\centering
\begin{tabular}{cc}
\hline 
Normalized mean-square error (NRMSE) & Normalized average absolute error (NAAE) \Tstrutb\Bstrutb\\
\hline
${\rm NRMSE}=\dis\sum_{i=1}^K\left(\hat{{\upsilon}}^{(i)}-{\upsilon}^{(i)}\right)^2 \bigg/ \dis\sum_{i=1}^K\left(\bar{\bm{\Upsilon}}-{\upsilon}^{(i)}\right)^2$
&
${\rm NAAE}=\left(K \mbf{\sigma}_{\bm{\Upsilon}}\right)^{-1} \, \dis\sum_{i=1}^K\left|\hat{{\upsilon}}^{(i)}-{\upsilon}^{(i)}\right|$ \Tstrut\Bstrut\\
\hline
Coefficient of Determination ($R^2$) & Normalized maximum absolute error (NMAE) \Tstrutb\Bstrutb\\
\hline
$R^2=\sigma^2_{\bm{\Upsilon}\hat{\bm{\Upsilon}}} \bigg/ \sigma^2_{\bm{\Upsilon}}\sigma^2_{\hat{\bm{\Upsilon}}}$
&
${\rm NMAE}=\left(K \mbf{\sigma}_{\bm{\Upsilon}}\right)^{-1} \,  \max_{i=1}^K\left|\hat{{\upsilon}}^{(i)}-{\upsilon}^{(i)}\right|$\Tstrut\Bstrut\\
\hline
\end{tabular}
\label{errormetrics}
\end{table}

In addition to the error metrics shown in Table \ref{error_metrics}, and to verify the whole rate of convergence of the proposed  model to the unknown function on  untried points,  we are interested in bounding the {\it maximum PCK-predictive error} over the domain ${\mathcal D}_{\textbf{x}}\subset \RR^M$:

\begin{equation}\label{unif}
\displaystyle\sup_{\textbf{x}\in {\mathcal D}_{\textbf{x}}}|{\mathcal M}(\textbf{x})-\hat{\mathcal M}_{PCK}(\textbf{x})|,\end{equation}

\noindent  where

$$\begin{array}{c}\hat{\mathcal M}_{PCK}(\textbf{x})=\dis\sum_{j\in{\mathcal J}}\chi_{{\mathcal  D}_{j}}({\textbf{x}})\hat{\mathcal  M}^{j}_{PCK} ({\textbf{x}})=\\\dis\sum_{j\in{\mathcal J}}\chi_{{\mathcal  D}_{j}}({\textbf{x}})\left[
{\textbf{r}^j}^\textrm{T}(\textbf{x}){\textbf{R}^j}^{-1}\bm{y}-\left( {\bm{\Theta}^j}^\textrm{T}{\textbf{R}^j}^{-1}{\textbf{r}^j}(\textbf{x})-{\mathbf{\Psi}^j}(\textbf{x})\right)^\textrm{T}\left({\bm{\Theta}^j}^{\textrm{T}}{\textbf{R}^j}^{-1} {\bm{\Theta}^j} \right)^{-1}{\bm{\Theta}^j}^{\textrm{T}}{\textbf{R}^j}^{-1}{\bm y}\right],
\end{array}
$$ 

\noindent is the best linear unbiased predictor (BLUP) of the model ${\mathcal M}$ response at any untried point $\textbf{x}\in {\mathcal D}_{\textbf{x}}$, with $\textbf{r}^j(\textbf{x}) = [R^j(|\textbf{x}-\textbf{x}^{(1)}|),\ldots R^j(|\textbf{x}-\textbf{x}^{(N_j)}|)]^\textrm{T}$  the vector of correlations between the design sites ${\Xi}_j=\left\{\textbf{x}^{(1)},\ldots,\textbf{x}^{(N_j)}\right\}\subset{\mathcal D}_j$ and $\textbf{x}$, and $R^j$ the selected correlation function particularized in the $j$-th subregion. Note that the uniform bound in Eq.~(\ref{unif}) covers the worst case for the prediction error of the PCK model.

It has been reported in the literature~\cite{Rit-2000,Wan-2020} that the prediction error of the universal Kriging converges to zero under uniform metric. Adapting Theorem 2 in \cite{Wan-2020} to the multielement PCE-Kriging model proposed in this work, the prediction error can be stated to satisfy:

\begin{equation}\label{error}\begin{array}{c}
 \mathbb{E}\left[\displaystyle\sup_{\mathbf{x} \, \in \, {\cal D}_{\mathbf{x}}}|{\mathcal M}(\mathbf{x})-\hat{\mathcal M}_{PCK}(\mathbf{x})| \right]\leq\\
 {\mathcal J}\cdot\displaystyle\max_{j\in{\mathcal J}}\mathbb{E}\left[\displaystyle\sup_{\mathbf{x} \, \in \, {\cal D}_{j}}|{\mathcal M}(\mathbf{x})-\hat{\mathcal M}_{PCK}(\mathbf{x})| \right]={\mathcal O}\left({\mathcal J}P_{{\Xi_j}} \left( {\mathcal P^j}A+\log^{\frac{1}{2}}P_{{\Xi_j}}^{-1} \right) \right),\end{array}
\end{equation}

\noindent where ${\mathcal J}$ is the number of subdomains, ${\mathcal P}^j=\textrm{card}\left(\mathcal{A}^{M,p,q}\right)$, and $A$ is a constant depending on  the eigenvalues of ${\mathbf{\Theta}}_j$. Term $P_{\Xi_j}(\textbf{x})$ denotes the {\it power function} given by $P^2_{\Xi_j}(\textbf{x}):=1-{\textbf{r}^j}^\textrm{T}(\textbf{x}){\textbf{R}^j}^{-1}{\textbf{r}}^j(\textbf{x})$, and $P_{\Xi}:=\sup_{\textbf{x} \, \in \, {\mathcal D}_\textbf{x}}P_{\Xi_j}(\textbf{x})$ is the supremum of the pointwise predictive standard deviation. It is thus reasonable to look for EDs  minimizing $P_{\Xi_j}$. Note that the rate of convergence in Eq.~(\ref{error}) is a deterministic function dependent on the experimental design ${{\Xi}_j}$ and decreasing with $P_{{\Xi_j}}$. In fact, when $N_j=$card$({\Xi_j})$ increases, $P_{{\Xi_j}}$  tends to zero and so does the multielement PCK prediction error under the uniform metric in Eq.~(\ref{error}).

On the other hand, the algorithm for finding an optimizer of Eq.~({\ref{ML}}) is an iterative process involving the calculation of the inverse and determinant of a large $N \times N$ covariance matrix ${R}_{ij}= R\left(\left| \textbf{x}^{(i)}-\textbf{x}^{(j)} \right|;\hat{\bm{\theta}} \right)$. Thus, the computational effort to obtain the solution may become impractical for large numbers $N$ of training data points in $\Xi$. Note that the PCK model requires ${\mathcal O}(N^3)$ operations and has a memory complexity of the order of $N^2$ \cite{Kon-2019}. In this light, the splitting technique presented in Section~\ref{multi} leads to substantial reductions in the computational effort. Specifically, taking $N_j={\rm card}({\Xi}_j)$, with $N_j\ll N$, ${\mathcal J}\ll N_j$, the algorithm effort and the memory storage reduces to ${\mathcal J}\cdot{\mathcal O}(N_j^3)\sim {\mathcal O}(N_j^3)\ll{\mathcal O}(N^3)$ and ${\mathcal J}\cdot{\mathcal O}(N_j^2)\sim {\mathcal O}(N_j^2)\ll {\mathcal O}(N^2)$, respectively. On the other hand, the optimal order of the polynomials in the PCE is automatically identified by the LAR algorithm. It is reported in reference~\cite{Efr-2004} that the LAR algorithm with $M$ variables requires ${\mathcal O}(M^3 + N_jM^2)$ computations in any subdomain ${\mathcal D}_j$. Therefore, in our case where $M\ll N_j$, it follows that $M^3<N_j M^2$ and, thus, ${\mathcal O}(N_jM^2)\sim  {\mathcal O}(N_j)$. Hence, the computational complexity of PCE when inserted as the trend term is marginal with respect to the overall construction of the Kriging model, thereby we can deduce that the efficiency of the proposed PCE-Kriging metamodel is ${\mathcal O}(N_j^3) \ll {\mathcal O}(N^3)$.


\subsection{Bayesian parameter inference via MCMC}\label{MCMCapproach}

In the Bayesian inference framework, model parameters $\bm{\theta}$ are conceived as a random variable with a certain posterior PDF $\pi$ described by Bayes' theorem:

\beq\label{bayes}
\pi\left(\bm{\theta}|\bm{y}\right) = \frac{p\left(\bm{y}|\bm{\theta}\right)p\left(\bm{\theta}\right)}{p\left(\bm{y}\right)}, \quad p\left(\bm{y}\right)=\int_\Omega p\left(\bm{y}|\bm{\theta}\right)p\left(\bm{\theta}\right) \textrm{d}\bm{\theta},
\eeq

\noindent~where $p(\bm{y}|\bm{\theta})={\cal L} (\bm{\theta})$ denotes the likelihood function, $p(\bm{\theta})$ the prior distribution of the model parameters, and $p(\bm{y}|{\cal M})$ a normalizing constant, also called evidence. In the context of this work, $\bm{y}$ and ${\bm \theta}$ represent a set of $n$ experimental observations and the model parameters of the metamodel to be calibrated, respectively. Errors $\bm{\eps}$ between the experiment and the predictions of the surrogate model are assumed to be normally distributed with zero mean and standard deviation $\sigma_{\bm{\eps}}$, that is $\bm{y} =\widehat{{\cal M}}(\bm{\theta})+\bm{\eps}$ with $\bm{\eps}\sim \mathcal{N}(0, {\sigma_{\bm{\eps}} \bm{I}})$. Then, the likelihood function ${\cal L} (\bm{\theta})$ can be expressed as: 

\beq\label{like}
{\cal L} (\bm{\theta})=\frac{\exp\left( -\dis\frac{1}{2\sigma_{\bm{\eps}}^2}\dis\sum_{i=1}^n|y_i-\widehat{{\cal  M}}(\bm{\theta})|^2\right)}{\sqrt{2\pi} \sigma_{\bm{\eps}}}.
\eeq

Obtaining $\pi$ from Eq.~(\ref{bayes}) in analytical closed-form is infeasible in most practical applications, being MCMC methods the most popular approach to numerically characterize the PDF of the model parameters. This approach allows one to draw samples from $\pi$ without computing the model evidence, which is independent from the model parameters $\bm{\theta}$. In this work, the DRAM algorithm developed by Haario \textsl{et al}.~\cite{Haa-2006} is implemented. This approach combines delayed rejection (DR)~\cite{Mir-2001} and adaptive Metropolis (AM)~\cite{Haa-2001}, which enhances the sampling efficiency of the sampling and enables the identification of multi-modal PDFs.  Given a set of observed data samples in vector \textbf{d}, the working principle of the DRAM approach can be outlined as follows:

\begin{enumerate}
\item Initialize the parameter set $\bm{\theta}_{c} = \bm{\theta}_0$ and the number $T$ of desired samples. Set an initial point from the parameter space and the covariance of the proposal distribution $\Sigma_{p}=\Sigma_{0}$. The proposal distribution is chosen as a multivariate Gaussian distribution with mean $\bm{\theta}_{c}$ and covariance matrix $\Sigma_{p}$. Select the initial non-adaptation period $n_o$ and set $i=1$.

\item Propose a new parameter value $\bm{\theta}_{p,1}$ by sampling from a proposal PDF $S_{1}(\bm{\theta},\bm{\theta}_{c})$. Accept $\bm{\theta}_{p,1}$ with probability:
\begin{equation}\label{Bay9}
\alpha_1\left(\bm{\theta}_{c},\bm{\theta}_{p,1}\right) = \min \left( 1,\frac{p\left(\left.\textbf{d} \, \right|\bm{\theta}_{p,1}\right)S_{1}(\bm{\theta}_{p,1},\bm{\theta}_{c})}{p\left(\left.\textbf{d} \, \right|\bm{\theta}_{c}\right)S_{1}(\bm{\theta}_{c},\bm{\theta}_{p,1})}\right),
\end{equation}
\noindent and go to step (4). If rejected, propose a second stage move in step (3).

\item Propose a second stage move $\bm{\theta}_{p,2}$ sampling from $S_2(\bm{\theta},\bm{\theta}_{p,1},\bm{\theta}_{c})$. This second stage proposal depends not only on the current position of the chain but also on the candidate that has just been proposed and rejected. Accept {\color{red} or} reject $\bm{\theta}_{p,2}$ by setting:
\begin{equation}\label{Bay11}
\bm{\theta}_{i} = \begin{cases}
\bm{\theta}_{p,2}, & \textrm{with probability} \, \alpha_2\left(\bm{\theta}_{c},\bm{\theta}_{p,1},\bm{\theta}_{p,2}\right), \\
\bm{\theta}_{c}, & \textrm{with probability} \, 1-\alpha_2\left(\bm{\theta}_{c},\bm{\theta}_{p,1},\bm{\theta}_{p,2}\right), \\
\end{cases}
\end{equation}

\noindent with

\begin{equation}\label{Bay10}
\alpha_2\left(\bm{\theta}_{c},\bm{\theta}_{p,1},\bm{\theta}_{p,2}\right) = \min \left\{ 1,\frac{p\left(\left.\textbf{d} \, \right|\bm{\theta}_{p,2}\right)S_1\left(\bm{\theta}_{p,2},\bm{\theta}_{p,1}\right)S_2\left(\bm{\theta}_{p,2},\bm{\theta}_{p,1},\bm{\theta}_{c}\right)\left[1-\alpha_1\left(\bm{\theta}_{p,2},\bm{\theta}_{p,1}\right)\right]}{p\left(\left.\textbf{d} \, \right|\bm{\theta}_{c}\right)S_1\left(\bm{\theta}_{c},\bm{\theta}_{p,1}\right)S_2\left(\bm{\theta}_{c},\bm{\theta}_{p,1},\bm{\theta}_{p,1}\right)\left[1-\alpha_1\left(\bm{\theta}_{c},\bm{\theta}_{p,1}\right) \right]}\right\}.
\end{equation}

\item Update the covariance matrix $\Sigma_{p}$ as:
\begin{equation}\label{Bay15}
\Sigma_{p} =         \begin{cases}
\Sigma_{0} & i \leq n_0 \\
s_d \textrm{cov}\left(\bm{\theta}_1,\ldots,\bm{\theta}_i\right) & i > n_0 \\
\end{cases}
\end{equation}
\noindent with $s_d$ a scaling parameter. Following \cite{Ste-2021}, $s_d=2.4^2/d$, with $d$ being the number of fitting parameters, is recommended as a good default value in most applications.
			
\item Go to step 2, until the desired number of samples $T$ are obtained.
\end{enumerate}


\section{Numerical results and discussion}\label{Sect3}

This section presents two application case studies to demonstrate the effectiveness of the proposed surrogate model-based Bayesian parameter estimation. These include a two-dimensional benchmark function and the PDE for TDS testing of hydrogen desorption in metals. The previous formulation has been implemented in Matlab environment, and all the numerical tests have carried out in a computer Intel(R) Core(TM) i9-10900X CPU$@$3.70 GHz with 64 GB of RAM memory. In the remainder of this section, for simplicity in the notation, the predictions of the PCK metamodels $\hat{\cal M}_{PCK}$ are noted as $\hat{\cal M}$. A $q$-norm value of 0.95 and Legendre polynomials of orders ranging from 2 to 6 are selected to build the PCEs in all the analyses hereafter. For the generation of the EDs, the number of random Monte Carlo candidate samples in the MIPT algorithm introduced in Section~\ref{PCK} is set to 25\,000.


\subsection{Two-dimensional Drop-Wave function}

This first case study investigates the Drop-Wave function, also known as the Salomon's function~\cite{Sal-1996}, given by $f:{\mathcal D}_{\tb x}=[-10,\,10]^2\in\RR^2\to\RR$:

\beq\label{salomonfunction}
f(x_1,x_2)=1-{\rm cos}\left(2\pi\sqrt{x_1^2+x_2^2}\right)+0.1\sqrt{x_1^2+x_2^2}.
\eeq

\begin{figure}[H]
\centering
\includegraphics[scale = 1]{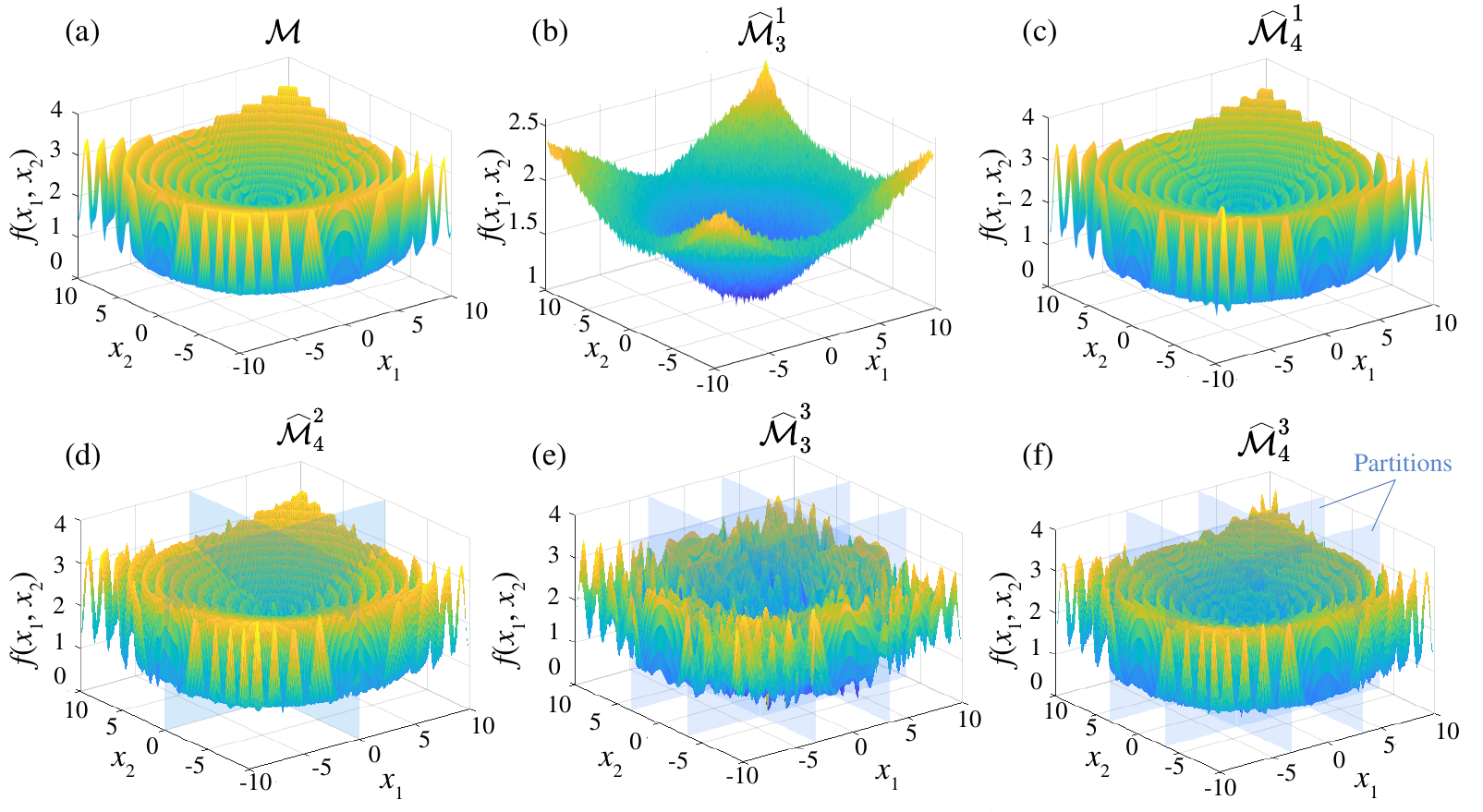}
\caption{Exact response surface of the Drop-Wave function (a), and predictions by surrogate models $\hat{\cal M}^3_1$ (b), $\hat{\cal M}_4^1$ (c), $\hat{\cal M}_4^{2}$ (d), $\hat{\cal M}^3_{3}$ (e) and $\hat{\cal M}_4^{3}$ (f).}
\label{Contour_Drop_Wave}
\end{figure}

This function is commonly used for benchmarking optimization algorithms. Owing to its highly non-linear character, the Drop-Wave function represents an ideal case study to validate the proposed multi-element PCK metamodel. Note that the surrogate modelling of this function using conventional techniques is extremely challenging given its fast-varying gradients and irregular response as observed in Fig.~\ref{Contour_Drop_Wave} (a). Following Section~\ref{multi}, four experimental design sets $ED_i\subset {\mathcal D}_{\tb x}$, $i=1,\ldots,4$ containing 360, 720, 1440 and 2880 samples have been defined. In addition, three different number of domain partition schemes ${\cal P}^j$, $j=1,\ldots,3$, have been considered. These include ${\cal P}^1=\left[-10,\,10\right]^2,$ ${\cal P}^2=\left(\left[-10,\,0\right)\cup \left[0,\,10\right]\right)^2$, and ${\cal P}^3=\left(\left[-10,\, \frac{-10}{3}\right)\cup \left[\frac{-10}{3},\, \frac{10}{3}\right)\cup \left[\frac{10}{3},\,10\right]\right)^2$, leading to a total of two, four and nine sub-domains, respectively. The number of samples has been chosen with the aim of obtaining a wide range of errors to correctly identify the convergence of the prediction error. For instance, if one takes the $R^2$ error metric, note that the constructed metamodels exhibit a wide range of $R^2$ values from $0.011$ to $0.999$. Additionally, the predictions by a previously reported multielement approach, the Stochastic Spectral Embedding (SSE) model proposed by Sudret and Marelli~\cite{Mar-2021}, are also presented as a reference solution. The SSE model is a PCE-based technique consisting of constructing a sequence of residual spectral expansions of the target model in subdomains of the input space. The implementation included in the UQLab software~\cite{UQL-2017} has been used to carry out the analyses. Four different surrogate models have been built, one for each considered ED. As parameters, a $q$-norm value of $0.95$, polynomials ranging from degree $2$ to $10$, and a minimum size of points per region equal to the size(ED)/$120$ have been selected. To sample the ED, the sequential experimental design based on the LHS implemented in UQLab has been chosen.

Hence, a total of sixteen surrogate models have been constructed. For ease in the discussion, the PCK surrogate models are specified with sub- and super-indexes denoting the size of the ED and the number of partitions, respectively, as reported in Table~\ref{12modelos}. All the surrogate models have been validated using a reasonably large VS of 20000 samples, and the accuracy of the models has been evaluated through the accuracy metrics reported in Table~\ref{metricsalomon}. The computational times involved in the construction $t_c$ of the surrogate models, as well as their evaluation times $t_e$ and $t_e$(VS) for a single point and the full VS have been also computed and collected to compare their effectiveness. The comparison of the exact response surface of the Drop-Wave function and the predicted ones by five of the constructed surrogate models is depicted in Fig.~\ref{Contour_Drop_Wave}.
 
\begin{table}[H]
\newcommand\Tstrut{\rule{0pt}{0.3cm}}         
\newcommand\Bstrut{\rule[-0.15cm]{0pt}{0pt}}   
\footnotesize
\caption{PCK surrogate-models constructed for the Drop-Wave function considering increasing EDs ($ED_i$) with varying numbers of domain partitions (${\cal P}^j$) (VS of 20\,000 samples).}
\vspace{0.1cm}
\centering
\begin{tabular}{ccccc}
\hline 
No. of sub-domains & $ED_1$ & $ED_2$ & $ED_3$ &$ED_4$ \Tstrut\\
& (360 samples) & (720 samples) & (1440 samples) & (2880 samples)\Bstrut\\
\hline
 ${\cal P}^1$ (1 partition) & $\hat{\cal M}_1^{1}$ & $\hat{\cal M}_2^{1}$ & $\hat{\cal M}_3^{1}$ & $\hat{\cal M}_4^{1}$ \Tstrut\\ 
 ${\cal P}^2$ (4 partitions) & $\hat{\cal M}_1^{2}$ & $\hat{\cal M}_2^{2}$ & $\hat{\cal M}_3^{2}$ & $\hat{\cal M}_4^{2}$\\
 ${\cal P}^3$ (9 partitions) & $\hat{\cal M}_1^{3}$ & $\hat{\cal M}_{2}^{3}$ & $\hat{\cal M}_{3}^{3}$ & $\hat{\cal M}_{4}^{3}$\Bstrut\\
  SSE (reference) & $\hat{\cal M}_1^{SSE}$ & $\hat{\cal M}_{2}^{SSE}$ & $\hat{\cal M}_{3}^{SSE}$ & $\hat{\cal M}_{4}^{SSE}$\Bstrut\\
\hline
\end{tabular}
\label{12modelos}
\end{table}

\begin{table}[H]
\newcommand\Tstrut{\rule{0pt}{0.3cm}}         
\newcommand\Bstrut{\rule[-0.15cm]{0pt}{0pt}}   
\footnotesize
\caption{Accuracy and computational efficiency analysis of proposed surrogate models applied to the Drop-Wave function. Terms $t_c$, $t_e$ and $t_e$(VS) denote the time of construction, average point evaluation, and evaluation on the full validation set, respectively.}
\vspace{0.1cm}
\centering
\begin{tabular}{cccccccc}
\hline 
Model & NAAE & NMAE & NRMSE & $R^2$ & $t_c$ [s] & $t_e$ [s] & $t_e$(VS) [s]\Tstrut\Bstrut\\
\hline
$\hat{\cal M}^1_{1}$&  1.030E+0             &  1.736E-4             &  1.589E+0             &     0.024           &      34.1      &      5.069E-5\ & 1.013\\ 
$\hat{\cal M}^1_{2}$&  8.340E-1         &  1.001E-4            &   8.701E-1            &  0.134              &  151.3           & 2.014E-4  & 4.028   \\
$\hat{\cal M}^1_{3}$&  8.321E-1         &   8.159E-5             &  8.627E-1             & 0.139               & 661.8          & 9.301E-4  & 18.601    \\
$\hat{\cal M}^1_{4}$&   2.876E-3          & 2.707E-5              & 2.651E-4             & 0.999              & 2905.7          &5.078E-3  & 101.540    \\
$\hat{\cal M}^2_{1}$&   9.162E-1          & 1.399E-4              & 12.080E-1            &      0.043               &  20.3          &  6.840E-6   & 0.138   \\
$\hat{\cal M}^2_{2}$&   8.978E-1         &  1.412E-4              & 11.980E-1           &       0.823                & 44.4           & 1.996E-5  & 0.399    \\
$\hat{\cal M}^2_{3}$&   5.448E-1        &   1.355E-4              & 4.974E-1             & 0.551          &      131.8           & 5.656E-5  & 1.131    \\
$\hat{\cal M}^2_{4}$&   2.256E-2         &  3.976E-5              & 2.559E-3             & 0.997              &  594.7          &  2.066E-4  & 4.132    \\
$\hat{\cal M}^3_{1}$&   8.757E-1         &  2.044E-4             &  11.230E-1              &    0.079               &  28.9          &  3.590E-6   & 0.071   \\
$\hat{\cal M}^{3}_{2}$& 7.765E-1       &   1.231E-4            &   8.622E-1           &   0.193               &  41.8         &  6.875E-6  & {0.138}    \\
$\hat{\cal M}^{3}_{3}$&  4.404E-1      &   1.297E-4           &    3.551E-1          &    0.645               &  86.4           & 1.513E-5  & {0.303}    \\
$\hat{\cal M}^{3}_{4}$&   4.345E-2     &   4.667E-5            &  6.712E-3             & 0.993               & 244.7 & 4.750E-5 & {0.950}\Bstrut\\
{$\hat{\cal M}^{SSE}_{1}$}&  1.249E+0      & 6.463E-4    & 2.684E+0  & 0.011  &  42.9  & 3.529E-5   &0.706  \\
{$\hat{\cal M}^{SSE}_{2}$}&  1.191E+0      & 1.302E-3    & 2.733E+0  & 0.024  &  33.9  & 3.401E-5   &0.681  \\
{$\hat{\cal M}^{SSE}_{3}$}&  8.838E-1      & 1.714E-3    & 1.535E+0  & 0.113  &  30.6  & 3.266E-5   &0.653  \\
{$\hat{\cal M}^{SSE}_{4}$}&  6.748E-1   & 5.694E-4   & 1.041E+0              & 0.298             & 32.5 & 3.101E-5 & 0.620\Bstrut\\
\hline
\end{tabular}
\label{metricsalomon}
\end{table}

\begin{figure}[H]
\centering
\includegraphics[width=\textwidth]{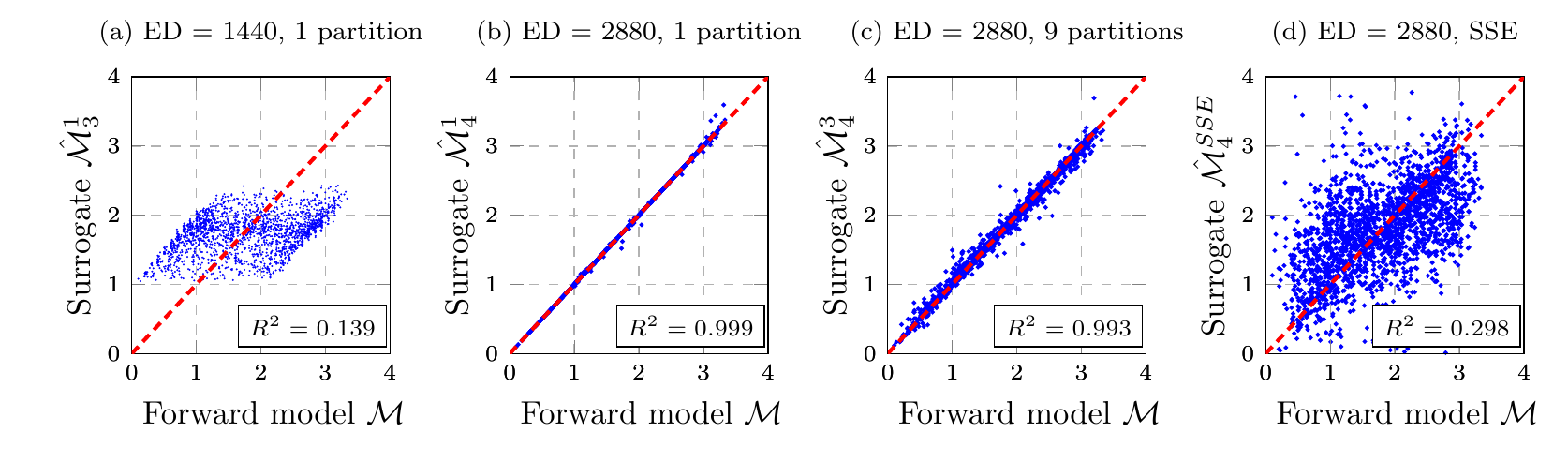}
\caption{Forward model evaluations versus surrogate model predictions for the Drop-Wave function. $\hat{\cal M}^1_3$  (a), $\hat{\cal M}^1_4$  (b),  $\hat{\cal M}^3_4$  (c) and $\hat{\cal M}^{SSE}_4$  (d) (VS of 20\,000 samples).}
\label{Scatter_Drop_Wave}
\end{figure}

\begin{figure}[H]
\centering
\includegraphics[scale=1]{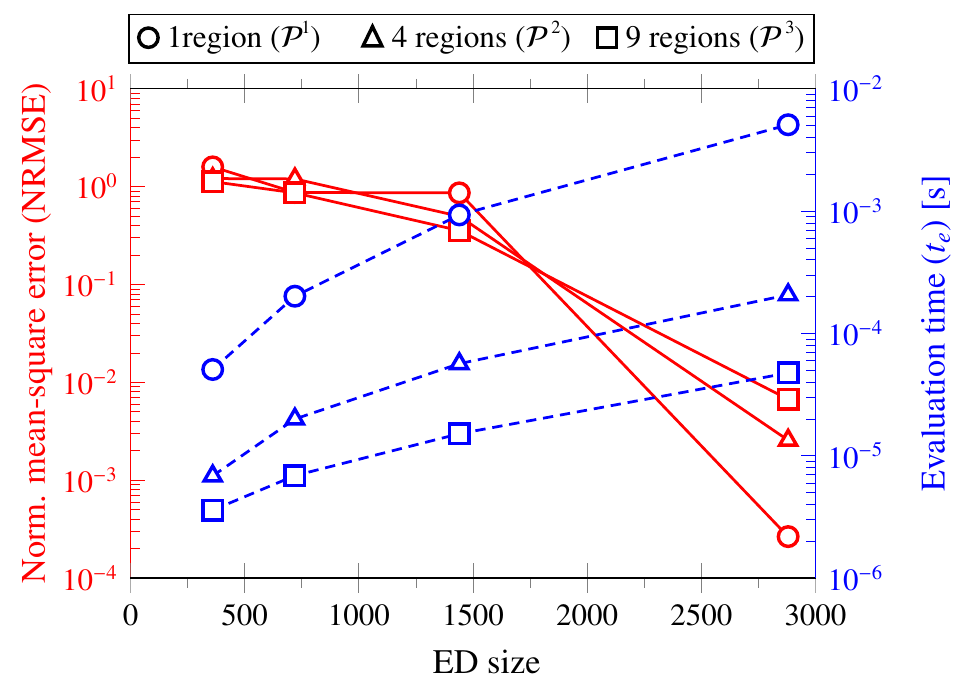}
\caption{Performance assessment of PCK surrogate models for the Drop-Wave function (VS of 20\,000 samples).}
\label{Perform_Drop_Wave}
\end{figure}

Figure~\ref{Scatter_Drop_Wave} shows the scatter plots of the forward model evaluated on the VS versus four metamodels, $\hat{\cal M}_3^1$ (a), $\hat{\cal M}_4^1$ (b), $\hat{\cal M}_4^3$ (c) and $\hat{\cal M}_4^{SSE}$ (d). The first two metamodels consider the whole design space, while the last two account for partition approaches. It is noted in this figure that the best approximations are found for $\hat{\cal M}_4^1$ (no partitions) and $\hat{\cal M}_4^3$ (9 partitions with EDs of 2880 samples). The predictions by these models exhibit low scatter around the diagonal line (perfect metamodel) with coefficients of determination R$^2$ very close to 1. The limited efficiency of the SSE model in this case study is evidenced by the large scatter of its predictions along the diagonal in Fig.~\ref{Scatter_Drop_Wave} (d). Interestingly, note that the predictions by $\hat{\cal M}_4^1$ slightly outperform those obtained with $\hat{\cal M}_4^{3}$, while higher numbers of partitions do not seem not to systematically improve the prediction accuracy. Nevertheless, the computational times involved in the construction and evaluation of $\hat{\cal M}_4^1$ are, respectively, about 10 and 100 times those required by $\hat{\cal M}_4^{3}$ (see Table~\ref{metricsalomon}). It is extracted from this analysis that the selection of the optimal surrogate model must be conducted by balancing the computational burden and the fitting accuracy. In this regard, Fig.~\ref{Perform_Drop_Wave} investigates the computational efficiency in terms of $t_e$ versus prediction accuracy (NRMSE) for all the considered surrogate models. In this figure, it is trivially observed that as the size of the ED increases, both computational time and accuracy of the metamodels increase. It is important to highlight that the consideration of higher number of subdomains leads to lower evaluation times for all the considered EDs. This is explained by the implementation of the LAR algorithm to extract optimal sets of polynomials in the PCE, and in particular, thanks to the reductions in the computational cost involved in the construction of the Kriging predictor (see Section \ref{MetricsAcc}). When inspected in a partition-wise fashion, the forward model exhibits a smoother behaviour, in such a way that the PCE requires less high-order polynomials to reproduce its behaviour. This results in more compact expansions, which also decreases the cost in the computation of the correlation matrix in the Kriging metamodel. On the other hand, note that the higher the order in the PCE, the larger the number of samples that are required in the ED to fit the expansion with accuracy. In this light, to reach a comparable accuracy to the one achieved by the Multi-Element PCK ($R^{2} >  0.99$) through SSE, it is necessary to increase the degree of the polynomial expansion up to 14 and to sample a ED with more than 45000 training points. These results strengthen the comparatively superior convergence rate of the proposed approach for this class of problems with highly non-linear spatial variability. As previously detailed in Section \ref{MetricsAcc}, the computational complexity of the proposed PCK model is ${\cal O}(N_j^3)$. This is due to the Cholesky decomposition of the correlation matrix $\textbf{R}$ in Eq.~(\ref{ML}). Therefore, as the ED increases, the computational cost involved in the determination of the stochastic hyper-parameters of the Kriging model and its evaluation rises dramatically. On the other hand, the dependence of the accuracy of the metamodel with the number of partitions is not so clear. It is noted that the accuracy of  metamodels trained with a larger number of subdomains is higher compared to those with no partitions for limited to moderate EDs. Nevertheless, when the size of the ED goes from moderate to large, the accuracy diminishes.

\subsection{Thermal Desorption Spectroscopy (TDS) of hydrogen in metals}

This last section reports the use of the proposed surrogate model for the Bayesian identification of hydrogen desorption and trapping characteristics in metallic materials. Hydrogen embrittlement (HE) refers to the loss of ductility and toughness of metallic alloys induced by hydrogen atoms deposited at lattice sites and micro-structural defects such as dislocations, grain boundaries or vacancies~\cite{Gangloff2003,Dwivedi2018}. Although this phenomenon has been extensively documented since the 19$^{\textrm{th}}$ century \cite{Johnson1875}, the growing trend towards a hydrogen-based economy as a means of mitigating $\textrm{CO}_2$ emissions and fossil fuel dependency has generated unprecedented interest on HE research. Micro-structural defects in metals act as `trap' sites, which sequester hydrogen and govern the susceptibility to HE \cite{AM2020,IJP2021}. Their characterization is thus of pivotal importance for the understanding of HE and the design of HE-resistant alloys, and this is generally achieved using TDS experiments \cite{Zafra2022}. The TDS test involves several stages~\cite{Cas-2002}: charging a sample with hydrogen, heating the sample at a fixed rate, and detecting the flux of desorbing hydrogen as a function of temperature by using a mass spectrometer. The hydrogen flow curve of desorbing hydrogen as a function of temperature defines the TDS spectrum, whose peaks can be associated with the presence of diverse micro-structural defects. Nonetheless, the formation of peaks in the TDS spectrum may be induced by the combined action of manifold hydrogen traps, being necessary to use simulation models and inverse calibration for their identification. Previous investigations on the modelling of the TDS test evidenced the existence of non-smooth relationships between the flux curves and the parameters characterizing micro-structural defects (see e.g.~\cite{Rai-2018}), making this application a formidable benchmark case study for the formulation presented in this work. In the remainder of this section, the PDE governing the hydrogen diffusion in materials tested by TDS is introduced in Section~\ref{theory}. The construction of the surrogate model and its performance evaluation is reported in Section~\ref{fittingTDS} and, finally, Section~\ref{BayesTDS} presents the Bayesian parameter identification results.

\subsubsection{TDS governing diffusion equation}\label{theory}

\begin{figure}[H]
\centering
\includegraphics[scale=1]{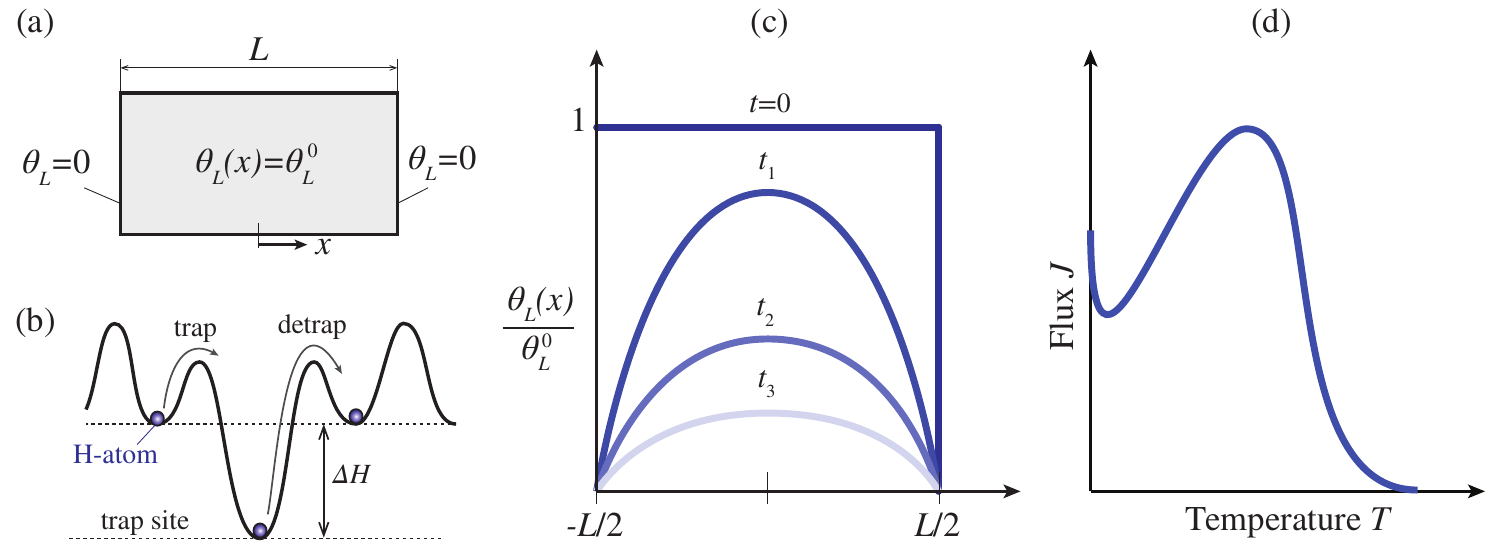}
\caption{(a) A schematic illustration of initial and boundary conditions in a TDS test. (b) Schematic definition of binding energy in a one-dimensional diffusion path. (c) Transient solution curves of the normalised lattice occupancy fraction $\theta_L/\theta^0_L$ at different times $t$ along the specimen's thickness. (d) A schematic of typical hydrogen desorption flux versus temperature curves obtained in a TDS test.}
\label{Scheme_TDS}
\end{figure}

Consider a one-dimensional specimen of length $L$ as sketched in Fig.~\ref{Scheme_TDS} (a). The specimen is subjected to increasing temperatures $T$, starting from $T_o$ and increasing at a constant heating rate $\phi$. Hydrogen atoms occupy normal intersticial lattice sites (NILS) and additionally can reside at trapping sites such as interfaces or dislocations. The kinetics of hydrogen trapping and detrapping in metals is commonly described with a two-level system as sketched in Fig.~\ref{Scheme_TDS} (b) for the case of a single trap. The potential landscape in this figure describes the diffusion path of hydrogen in metals, the trap binding energy $\Delta H$ being the difference between detrapping and trapping energies. Let us assume that the number of hydrogen traps in the specimen amounts to $N_t$. Then, let $C_L(x,t)$ and $C_{T,i}(x,t)$, $i=1,\ldots,N_t$, denote the hydrogen concentration in the lattice and in the $i$-th trap, respectively, with $x \in \left[-L/2,L/2\right]$ and $t$ respectively denoting space and time. On this basis, the Fickian diffusion equation needs to be enriched with source and sink terms as~\cite{Rai-2018}:

\beq\label{TDS_for1}
\frac{\partial C_L}{\partial t}+\sum_{i=1}^{N_t}\frac{\partial C_{T,i}}{\partial t}=D_L\frac{\partial^2 C_L}{\partial x^2},
 \eeq

\noindent with $D_L=D_o\exp\left(-Q/RT\right)$ being the lattice diffusion coefficient, which is expressed in terms of the lattice activation energy $Q$, diffusion pre-exponential factor $D_o$, and the universal gas constant $R$. It is convenient to introduce the lattice and trap occupancy fractions $\theta_L$ and $\theta_{T,i}$ $\left(\left\{ \theta_L, \theta_{T,i}\right\} \in \left[0,1\right]\right)$, respectively, by rewriting the corresponding concentrations in the form $C_L=\theta_L \, \beta \, N_L$ and $C_{T,i}=\theta_{T,i} \, \alpha \, N_{T,i}$. Here, $\beta$ is the number of NILS per unit volume, $\alpha$ is the number of atoms sites per trap, $N_L$ is the number of lattice atoms per unit volume, and $N_{T,i}$ is the number of trap sites per unit volume. Therefore, Eq.~(\ref{TDS_for1}) can be rewritten as:

\beq\label{TDS_for2}
\frac{\partial \theta_L}{\partial t}+\sum_{i=1}^{N_t} \left(\frac{\alpha N_{T,i}}{\beta N_L}\right) \frac{\partial \theta_{T,i}}{\partial t}=D_L\frac{\partial^2 \theta_L}{\partial x^2}.
\eeq

The PDE in Eq.~(\ref{TDS_for2}) needs to be complemented with trap kinetic equations. To this aim, the formulations by MacNabb and Foster~\cite{Nab-1963} and Oriani~\cite{Oriani1970} are commonly adopted. The latter represents a simplification of the former by assuming that a local equilibrium exists between the hydrogen atoms at the lattice sites and the $i$-th trap such that, for $\theta_L \ll 1$, 

\beq\label{TDS_for3}
\theta_{T,i}=\frac{K_i\theta_L}{\left(1+K_i\theta_L\right)} \, , 
\eeq

\noindent with $K_i$ being the local equilibrium constant for the $i$-th trap:

\beq\label{TDS_for4}
K_i = \exp\left\{ -\frac{\Delta H_i}{RT} \right\}.
\eeq

Introducing Eq.~(\ref{TDS_for3}) into (\ref{TDS_for2}), and following the non-dimensional formulation developed by Raina \textit{et al}.~\cite{Rai-2018}, the governing PDE describing hydrogen diffusion in the TDS test can be recast in a compact form as: 

\beq\label{TDS_for5}
\frac{\partial \overline{\theta}_L}{\partial \overline{t}}\left[1+\sum_{i=1}^{N_t}\frac{\overline{K}_i \, \overline{N}_i}{\left(1+\overline{K}_i \, \theta_L^0 \, \overline{\theta}_L\right)^2}\right]+\frac{\overline{\theta}_L}{\overline{T}^2}\sum_{i=1}^{N_t}\frac{\overline{K}_i \, \overline{N}_i\,\overline{\Delta H}_i \, \overline{\phi}}{\left(1+\overline{K}_i \, \theta_L^0 \, \overline{\theta}_L\right)^2} = \overline{D}_L\frac{\partial^2 \overline{\theta}_L}{\partial \overline{x}^2},
\eeq

\noindent with $\theta_L^0$ being the initial lattice occupancy. The non-dimensional variables employed are listed in Table~\ref{nondimTDS}. 

\begin{table}[H]
\newcommand\Tstrut{\rule{0pt}{0.3cm}}         
\newcommand\Bstrut{\rule[-0.15cm]{0pt}{0pt}}   
\footnotesize
\caption{Non-dimensional variables used in the hydrogen diffusion PDE employed for the TDS tests.}
\vspace{0.1cm}
\centering
\begin{tabular}{llll}
\hline 
Spatial coordinate & $\overline{x}=x/L$ & Lattice activation energy & $\overline{Q}=Q/\left( R  T_o \right)$\Tstrut\\ 
Time coordinate & $\overline{t}=tD_o/L^2$ & Trap binding energy & $\overline{\Delta H}_i=\Delta H_i/\left( R  T_o \right)$\\
Heating rate & $\overline{\phi}=\left( \phi L^2 \right)/\left( T_o D_o \right)$ & Lattice diffusion coefficient & $\overline{D}_L=D_L/D_o$\\
Trap density & $\overline{N}_i=\left( \alpha N_{T,i} \right)/\left( \beta N_L \right)$ & Local equilibrium constant & $\overline{K} = \exp\left\{ -\frac{\overline{\Delta H_i}}{\overline{T}} \right\}$\\
Temperature & $\overline{T}=T/T_o$ & Fractional lattice occupancy & $\overline{\theta}_L=\theta_L/\theta_L^o$\Bstrut\\
\hline
\end{tabular}
\label{nondimTDS}
\end{table}

The initial and boundary conditions of the PDE in Eq.~(\ref{TDS_for5}) are schematically presented in Fig.~\ref{Scheme_TDS} (a). At $t=0$, it is assumed an initial uniform lattice occupancy $\overline{\theta}_L\left(\overline{x},\overline{t}=0\right)=1$. Thereafter, the hydrogen lattice occupancy is assumed zero at the boundaries, that is $\overline{\theta}_L\left(\overline{x} = \pm 1/2,\overline{t}>0\right)=0$. As temperature raises, the lattice occupancy evolves spatially and temporally as sketched in Fig.~\ref{Scheme_TDS} (c), and the flux of hydrogen atoms $J(t)$ diffusing out at boundaries is measured as presented in Fig.~\ref{Scheme_TDS} (d). This flux can be obtained in non-dimensional terms after solving Eq.~(\ref{TDS_for5}) as~\cite{Rai-2018}:

\beq\label{TDS_for6}
\overline{J} = -\overline{D}_L \, \theta_L^o \, \frac{\partial \overline{\theta}_L}{\partial \overline{x}}. 
\eeq

Generally, the magnitudes of ${Q}$, $D_0$ and $\theta^0_L$ are known, and the heating rate ${\phi}$ is an input to the TDS system. Therefore, the TDS spectrum can be used to map the microstructural hydrogen traps, as characterised by their trap densities $(\overline{N}_i)$ and binding energies $(\overline{\Delta H}_i)$. These can be obtained for a given flux curve $\overline{J}$ by the inverse calibration of the PDE in Eq.~(\ref{TDS_for5}).

The surrogate modelling of the flux curves obtained after solving Eq.~(\ref{TDS_for5}) represents a formidable problem due to the strong nonlinearities of these curves. Specifically, depending upon the hydrogen trap configuration, several different regimes can be observed, as previously discussed by Raina \textit{et al}.~\cite{Rai-2018}. Specifically, their results for the case of metals containing a single trap showed that no peak flux is attained for low trap densities and binding energies. Alternatively, when a peak flux is found, those authors identified two distinct regimes (I and II) originated by two types of microstructural defects, referred to as shallow and deep traps. Shallow traps are characterized by large trap densities, and give origin to peak fluxes that are highly sensitive to both ${N}$ and $\Delta H$. On the other hand, deep traps are characterized by low trap densities, resulting in peak fluxes that are insensitive to the trap binding energy. The existence of these different regimes turns the construction of a surrogate model covering the whole domain of the traps into an notably challenging task. Note that a large number of high-order polynomials and a dense ED need to be included in the PCE to accurately represent the whole global behaviour of the hydrogen flux. Such large EDs may severely compromise the computational efficiency of the surrogate model since, as indicated above, the complexity of the Cholesky decomposition of the correlation matrix $\textbf{R}$ in Eq.~(\ref{ML}) is $\mathcal{O}\left(N^3\right)$. The metamodeling of TDS experiments thus represents an exceptional case study to justify the use of the domain partitioning approach presented in Section~\ref{multi}.

\subsubsection{Surrogate modelling of TDS flux curves for metals with two traps}\label{fittingTDS}

The PDE in Eq.~(\ref{TDS_for5}) is solved numerically by using the \textit{pdepe} solver in MATLAB. A space discretization of 201 elements along $x$ was found to provide mesh-independent results. To illustrate the behaviour of a ferritic steel sample, representative model parameters from reference~\cite{Rai-2018} have been adopted herein, including a lattice activation energy $Q=6.7$ kJmol$^{-1}$, diffusion pre-exponential factor $D_o=2 \times 10^{-7}$ m$^2$s$^{-1}$, heating rate $\overline{\phi} = 0.1$ and lattice density $N_L=8.46 \times 10^{28}$ atoms\,m$^{-3}$, with $\alpha=\beta=1$. The initial temperature and occupancy fraction are chosen as $T_o=293$ K and $\theta_L^0=10^{-6}$, respectively, and the thickness of the specimen is chosen as $L=5$ mm. The variation of trap binding energies and trap densities are selected as the physically meaningful ranges $-40 \leq \overline{\Delta H}_i \leq -10$ and $10^{-7} \leq \overline{N}_i \leq 10^{-2}$. In the present study, we limit to the modelling of metals with two hydrogen traps, i.e.~$N_t=2$. Therefore, in the surrogate modelling, temperature and the trap densities and binding energies are considered as input variables, which amounts to 5 design variables, i.e.~${\cal M}(\bm{x}) = \overline{J}$ with $\bm{x} = \left[\overline{T},\overline{\Delta H}_1,\overline{\Delta H}_2,\log(\overline{N}_1), \log(\overline{N}_2)\right]^\textrm{T} \subset \RR^5$. 

\begin{table}[H]
\newcommand\Tstrut{\rule{0pt}{0.3cm}}         
\newcommand\Bstrut{\rule[-0.15cm]{0pt}{0pt}}   
\footnotesize
\caption{Accuracy and computational efficiency analysis of PCK metamodels developed for the surrogate modelling of hydrogen diffusion flux curves obtained by TDS (VS=36\,000).}
\vspace{0.1cm}
\centering
\begin{tabular}{lcccc}
\hline 
& $\hat{\cal M}_{1}$ & $\hat{\cal M}_2$ & $\hat{\cal M}_3$ & $\hat{\cal M}_4$\Tstrut\Bstrut\\
& 72 subdomains & 72 subdomains & 108 subdomains & 108 subdomains\Tstrut\Bstrut\\
& ED = 72\,000 & ED = 108\,000 & ED = 72\,036 & ED = 108\,000\Tstrut\Bstrut\\
\hline
$t_e$ [ms] & 1.97 & 4.86 & 0.83 & 2.02\Tstrut\\
NAAE & 1.0637E-02 & 7.540E-03 & 1.115E-02 & 7.736E-03\Tstrut\\
NMAE & 2.517E-05 & 2.239E-05 & 2.363E-05 & 1.656E-05\\
NRMSE & 9.452E-04 & 5.259E-04 & 1.208E-03 & 5.229E-04\\
$R^2$ & 0.9991 & 0.9995 & 0.9988 & 0.9995\Bstrut\\
\hline
\end{tabular}
\label{metricTDS}
\end{table}

\begin{figure}[H]
\centering
\includegraphics[scale = 1]{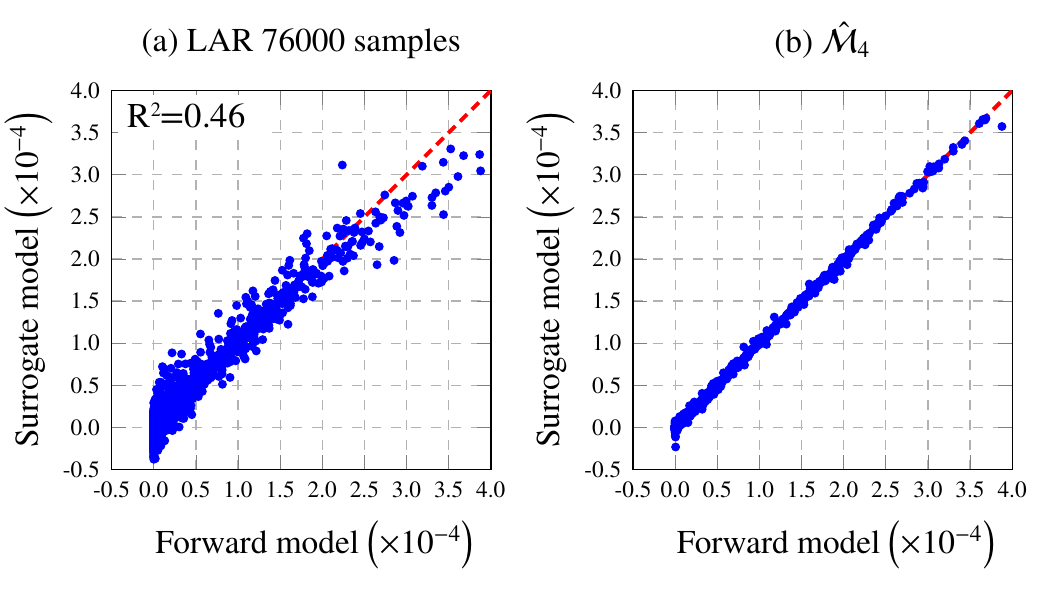}
\caption{Scatter plots of Hydrogen flux curves obtained by the forward solution of the PDF of the TDS test versus the predictions by standard LAR-PCE (a) and by the proposed multi-element PCK metamodel $\hat{\cal M}_4$ (b) (VS of 36\,000 samples).}
\label{scatter_TDS_proc}
\end{figure}

\begin{figure}[H]
\centering
\includegraphics[width=1.00\textwidth]{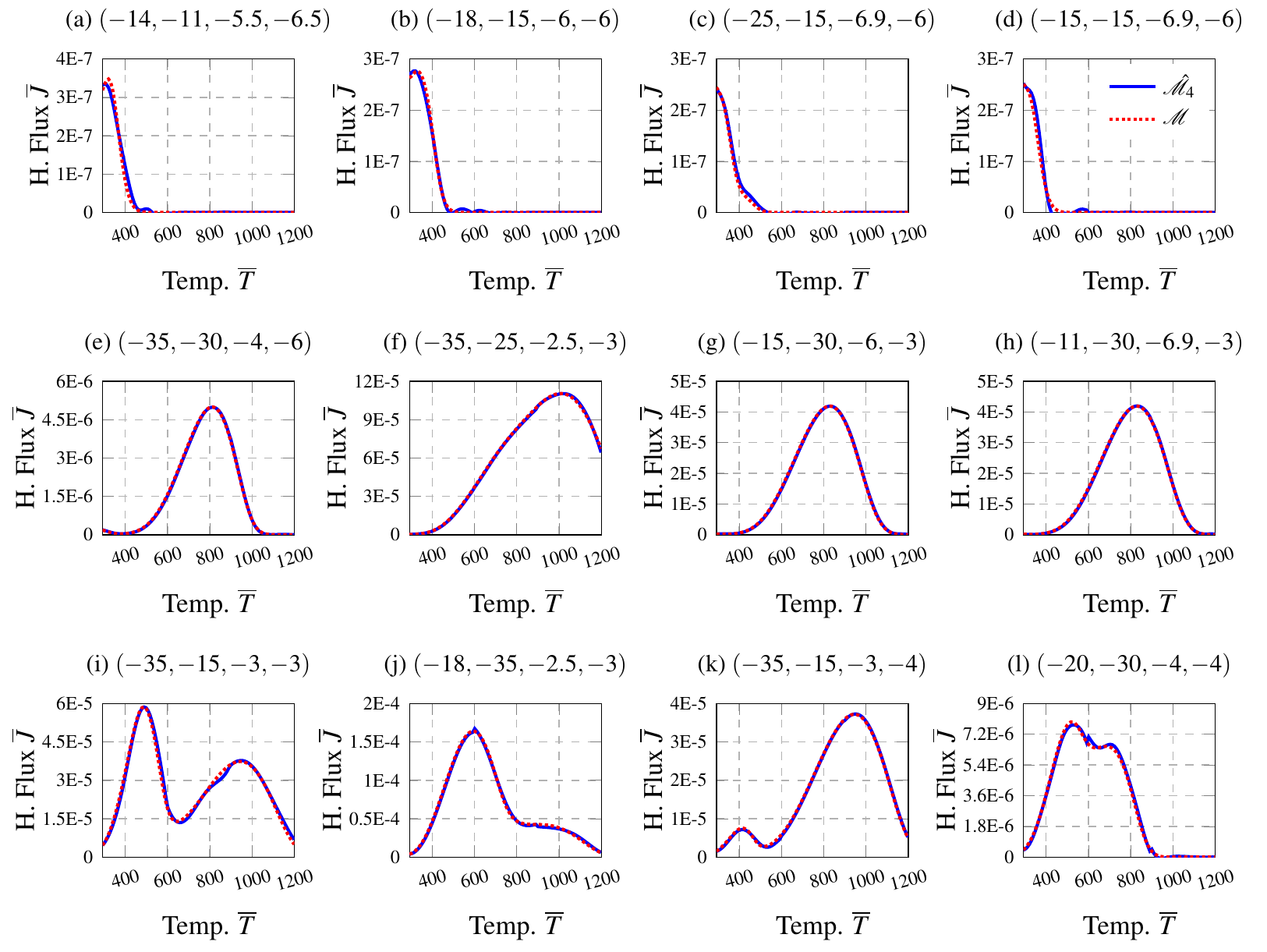}
\caption{Surrogate modelling of the Hydrogen flux curves obtained by TDS of metals with different values of trap binding energies and concentrations. Quantities in parenthesis represent the parameters of the traps $\left(\overline{\Delta H}_1,\overline{\Delta H}_1,\log(\overline{N}_1), \log(\overline{N}_2)\right)$. }
\label{flux_surrogate_versus_EDO}
\end{figure}

After some preliminary sensitivity analyses, two partitions of ${\cal D}_{\tb x}$ have been considered, namely ${\cal P}^1,\,{\cal P}^2$. Partition ${\cal P}^1$ has been defined by splitting the temperature $\overline{T}$ and activation energy domains ($\overline{\Delta H}_i$, $i=1,\,2$) in two, while three segments were considered for the partition of the domain of the trap densities ($\log(\overline{N_i})$, $i=1,\,2$). On the other hand, the partitions in ${\cal P}^2$ remain identical except for the temperature domain which is divided in three sub-domains. In order to define the optimal surrogate model, EDs of 1000 and 1500 sampling points per-subdomain have been considered for ${\cal P}^1$, while EDs of 667 and 1000 points per sub-domain have been defined for ${\cal P}^2$. This amounts to four different surrogate models labelled with $\hat{\cal M}_i$, $i=1,\ldots,4$. In order make a fair comparison between the different proposals, models $\hat{\cal M}_1$ and $\hat{\cal M}_3$ are trained with EDs of 72\,000 and 72\,036 samples respectively, while $\hat{\cal M}_2$ and $\hat{\cal M}_4$ are trained with 108\,000 samples. The comparison of the metamodels in terms of accuracy and computational efficiency is reported in Table~\ref{metricTDS} over a VS of 36\,000 samples. Similarly to the results in the previous case study, the consideration of domain partitioning leads to considerable computational time reductions and moderate reductions in prediction accuracy. Note that the evaluation time of the forward model is about 280 ms, so all the metamodels achieve reductions between 98.3\%-99.7\%. The computation time of the metamodel depends upon the size of the ED in each region, which explains why models $\hat{\cal M}_3$ and $\hat{\cal M}_2$ are the fastest and slowest ones, respectively. On the other hand, the accuracy of the metamodel increases as so does the size of the ED. Indeed, models $\hat{\cal M}_2$ and $\hat{\cal M}_4$ exhibit significantly lower errors compared to models $\hat{\cal M}_1$ and $\hat{\cal M}_3$. Therefore, in view of these results, $\hat{\cal M}_4$ provides a good trade-off between computational efficiency and accuracy, and it is selected in the subsequent Bayesian model parameter inference. To illustrate the effectiveness of the surrogate model in representing the different stages observed in the TDS test, Fig.~\ref{flux_surrogate_versus_EDO} shows the comparison of the forward model and the predictions by $\hat{\cal M}_4$ for a variety of combinations of traps, including the case of fluxes without peak, one single peak, and two peaks. It is observed that the proposed PCK model can accurately  reproduce all the different regimes observable in the TDS test. Only some minor errors are observed in the no flux regime, given the imposed limitation on the order of the polynomials in the PCE for the sake of computational efficiency. Finally, in order to highlight the superior performance of the proposed multi-element PCK metamodel, Fig.~\ref{scatter_TDS_proc} furnishes the comparison of the predictions by standard LAR-PCE (trained with 76\,000 samples) and $\hat{\cal M}_4$. These results clearly evidence the superior performance of the proposed approach with respect to LAR-PCE, whose predictions versus the forward model exhibits a large scatter around the diagonal line with a low coefficient of determination of $R^2=0.46$.  

\subsubsection{Bayesian inference of the trapping sites from a TDS experiment}\label{BayesTDS}

\begin{figure}[H]
\centering
\includegraphics[scale=1.0]{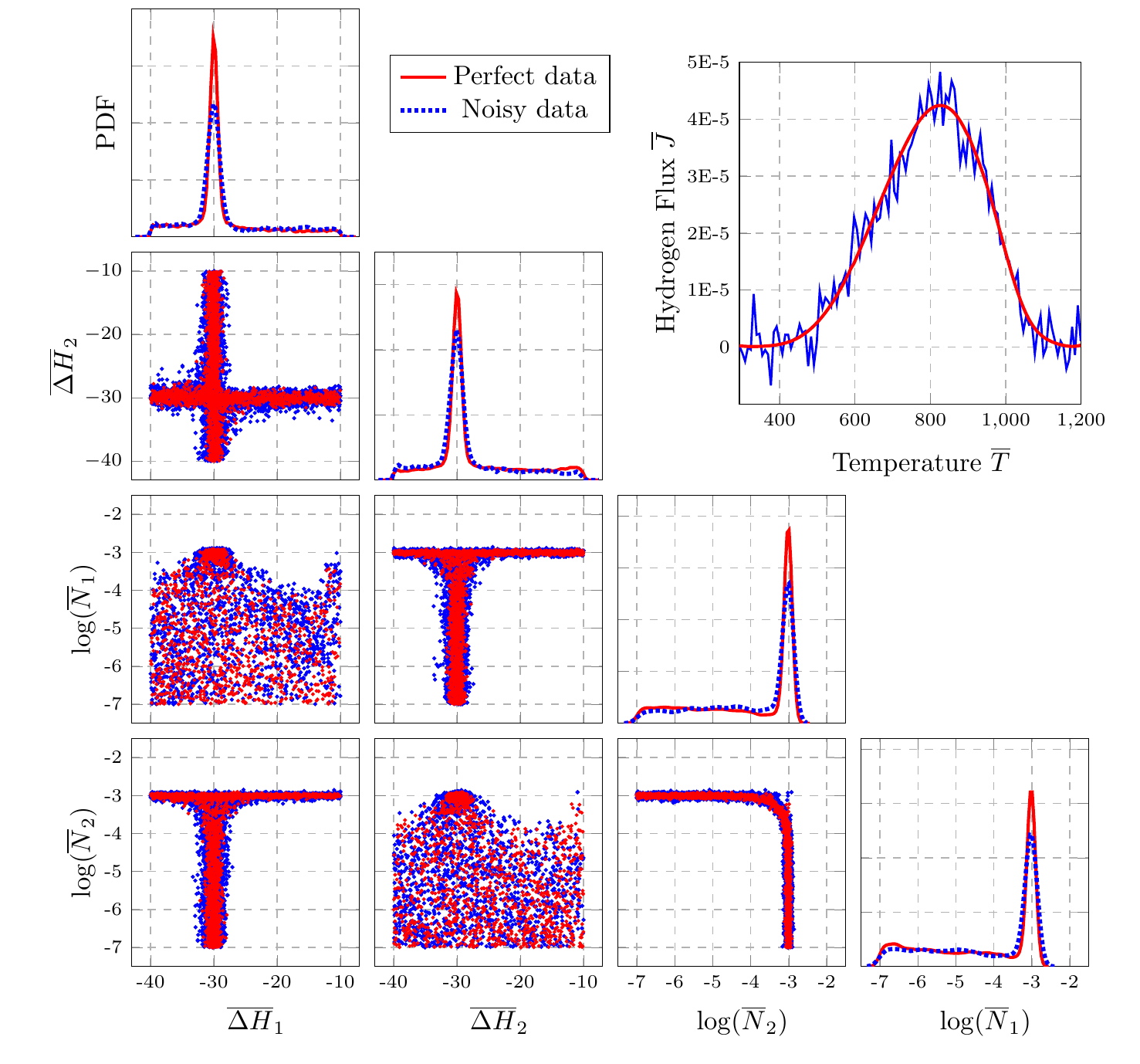}
\caption{Bayesian identification results of the trap parameters $\bm{\theta}=\left(\overline{\Delta H}_1,\overline{\Delta H}_2,\log(\overline{N}_1),\log(\overline{N}_2)\right)$ of experiment EII with noise unpolluted (red) and polluted data (blue).}
\label{PDF_case2}
\end{figure}

In this last subsection, the previous surrogate model $\hat{\cal M}_4$ is used to conduct Bayesian parameter inference following the MCMC algorithm in Section~\ref{MCMCapproach}. The trap binding energies and densities of the two trap system are chosen as the inference parameters $\bm{\theta}=\left(\overline{\Delta H}_1,\overline{\Delta H}_2,\log(\overline{N}_1),\log(\overline{N}_2)\right)$ in Eq.~(\ref{bayes}). With the purpose of assessing the performance of the implemented DRAM MCMC approach to infer the parameters of hydrogen traps covering the two different regions described in reference \cite{Rai-2018}, two different trap configurations are considered to generate synthetic experimental data from the forward model. A two-trap system (EI) with properties $\bm{\theta}=\left(-25, -35, -3, -2.5\right)$ is considered first. The second one (EII) instead is defined by $\bm{\theta}=\left(-15, -30,-6, -3\right)$. The flux curves obtained in EI and EII correspond to those previously shown in Figs.~\ref{flux_surrogate_versus_EDO} (f) and (g), respectively. In addition, to evaluate the sensitivity of the model parameter inference to the presence of noise pollution in the experiment, a second analysis of the EII experiment was performed after affecting the flux curve with a zero-mean Gaussian white noise with a standard deviation equal to 0.4 times the mean value of the unpolluted flux curve (note later in Fig.~\ref{PDF_case2} that such a noise level represents a considerably low signal-to-noise ratio). The experiment EI was defined to illustrate the potentials of the implemented DRAM MCMC to draw samples from a multi-modal distribution. Note that the PDE in Eq.~(\ref{TDS_for5}) does not differentiate the order of the traps, thereby the problem is ill-posed and the posterior distribution is expected to exhibit two modes corresponding to two symmetric solutions. Instead, the experiment EII was designed to account for a trap ($\overline{\Delta H}_1 = -15$, $\log(\overline{N}_1) = -6$) in the regime with no flux as identified by Raina and \textit{et al.}~\cite{Rai-2018}, while the second trap ($\overline{\Delta H}_2 = -30$, $\log(\overline{N}_2) = -3$) represents a deep trap. Therefore, the PDF in this case should be uni-modal.  

In the inference analyses, uninformative uniform priors $\mathcal{U}(-40,-10)$ and $\mathcal{U}(-7,-2)$ are selected for $\overline{\Delta H}_i$ and $\log(\overline{N}_i)$ ($i=1,\,2$), respectively. A total number of 200\,000 samples with a burning time of 50\,000 samples were drawn by the previously introduced Bayesian inference approach for EI. The sampling of the posterior PDF in Experiment EII was more challenging given its uni-modal nature with large regions of low probability, requiring up to 480\,000 samples with a burning period of 160\,000 samples to achieve convergence. Interestingly, this phenomenon attenuates when the flux curve is affected by noise, only requiring a chain of 120\,000 samples with a burning period of 30\,000 to attain convergence. This is expectable since the noise-induced lower probability concentration around the exact true solutions makes it easier for the chain to span from one solution to the symmetric one. The initial location state was defined as $\bm{\theta}_0 = \left(-25, -25, -4.5, -4.5 \right)$, while the prediction error was set to $\sigma_{\bm{\eps}} = 1E-9$ and $1.6E-5$ for the noise unpolluted and polluted cases, respectively. After some initial calibration by visual inspection of the chain traces, a diagonal covariance matrix with entries equal $\left(0.05\cdot \bm{\theta}_0\right)^2$ was initially defined for the Gaussian proposal. In the AM step the proposal distribution was scaled by a factor $s_d=2.4^2/d$ and the non-adaptation period $n_0$ was set to $500$ and $4000$ for the EI and EII experiments, respectively. On the other hand, in the DR step the proposal is scaled down by a factor of 0.2. 

\begin{figure}[H]
\centering
\includegraphics[scale=1.0]{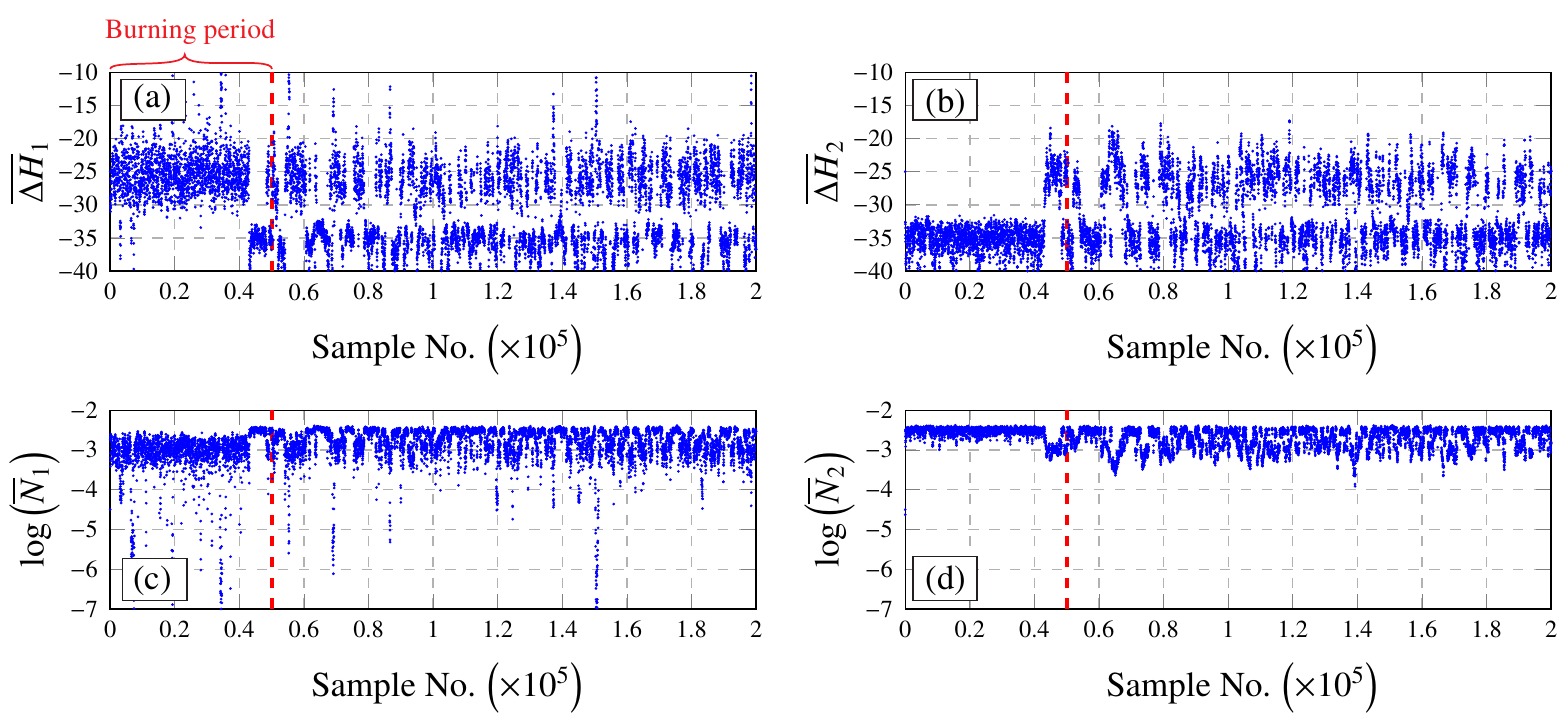}
\caption{Markov chains generated by DRAM MCMC of trap parameters $\overline{\Delta H}_1$, $\overline{\Delta H}_2$, $\log(\overline{N}_1)$, $\log(\overline{N}_2)$ for TDS Experiment EI.}
\label{Chain_case_I}
\end{figure}

\begin{figure}[H]
\centering
\includegraphics[scale=1.0]{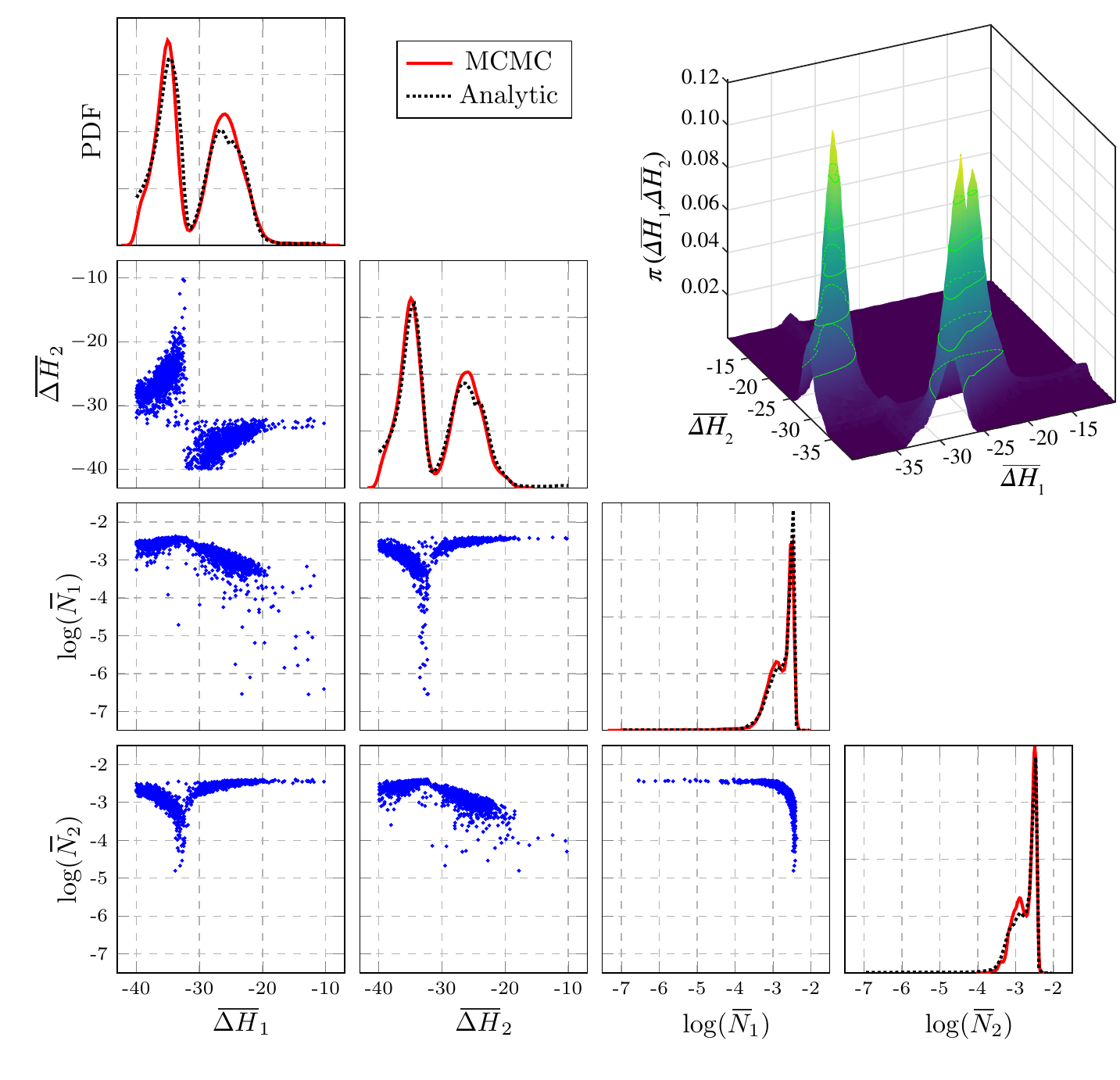}
\caption{Bayesian identification results of the trap parameters $\bm{\theta}=\left(\overline{\Delta H}_1,\overline{\Delta H}_2,\log(\overline{N}_1),\log(\overline{N}_2)\right)$ of TDS Experiment EI. The surface plot in the top right corner corresponds to the marginal PDF over $\left(\overline{\Delta H}_1,\overline{\Delta H}_2\right)$ obtained by numerical integration.}
\label{PDF_case1}
\end{figure}

\begin{figure}[H]
\centering
\includegraphics[scale=1.0]{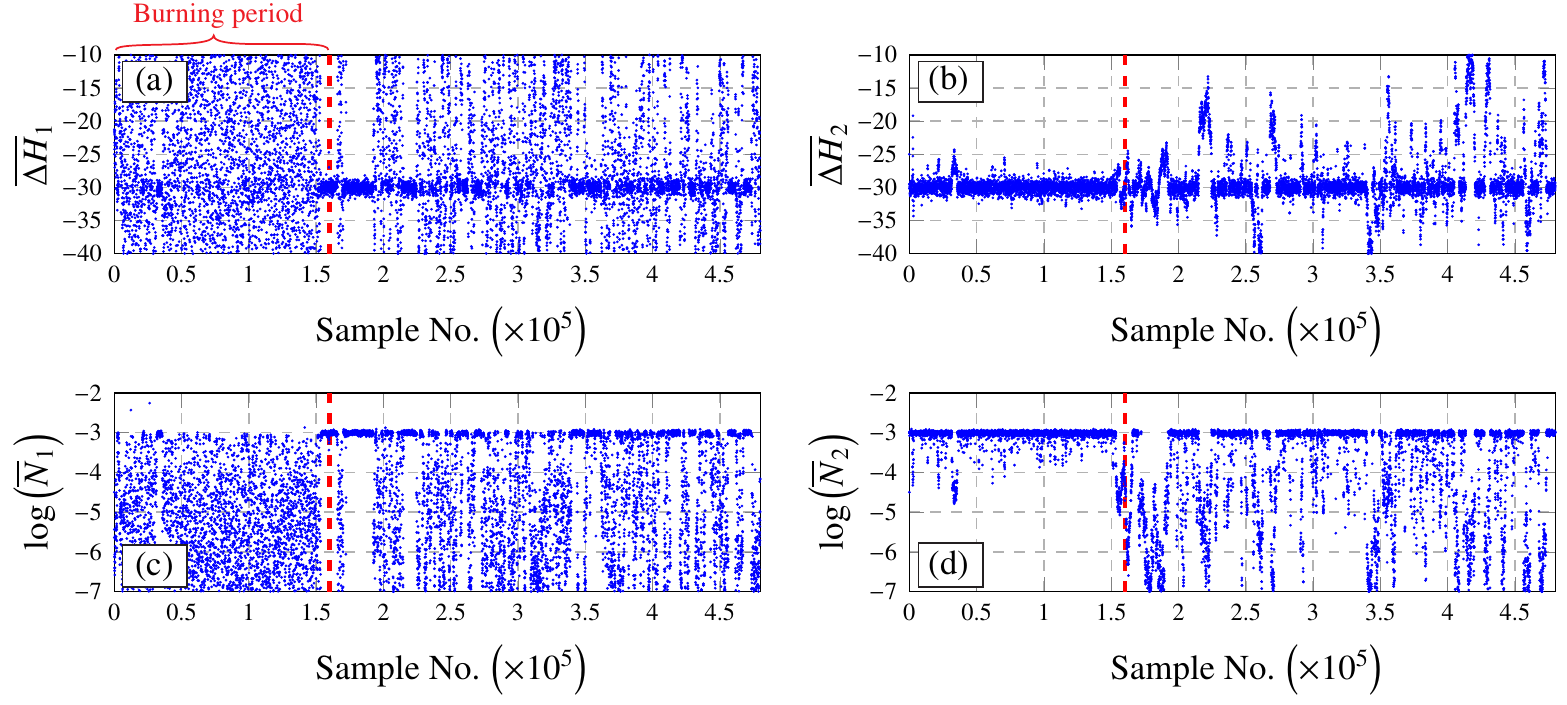}
\caption{Markov chains generated by DRAM MCMC of trap parameters $\overline{\Delta H}_1$, $\overline{\Delta H}_2$, $\log(\overline{N}_1)$, $\log(\overline{N}_2)$ for TDS Experiment EII under noise-free data conditions.}
\label{Chain_case_II}
\end{figure}

The Markov chain and the joint posterior PDF obtained for Experiment EI are presented in Figs.~\ref{Chain_case_I} and \ref{PDF_case1}, respectively. As anticipated, the problem is ill-posed and there exist two potential solutions, namely $\bm{\theta}=\left(-25, -35, -3, -2.5\right)$ and $\bm{\theta}=\left(-35, -25, -2.5, -3\right)$. This manifests in the marginal PDFs in Fig.~\ref{PDF_case1}. Specifically, the PDFs corresponding to parameters $\overline{\Delta H}_1$ and $\overline{\Delta H}_2$ have two identical modes at $-35$ and $-25$, and parameters $\log(\overline{N}_1)$ and $\log(\overline{N}_2)$  have two modes at $-3$ and $-2.5$. It is observed in Fig.~\ref{PDF_case1} that, indeed, the implemented DRAM algorithm is capable of exploring the two modes in the distribution, without getting stuck around one of them as it is usually the case when implementing standard MCMC methods. For validation purposes, the posterior PDF has been also computed by direct integration of the forward solution. To do so, the evidence of the model has been computed over a mesh of $60^4$ elements. This required forty five hours of parallel computation on ten cores, while the MCMC approach only required about four hours on a single core. The Highest Density Regions (HDRs) at the 80\% and 50\% level of both distributions are reported in Table~\ref{EI_HDR_table}. The close fittings between the exact marginal PDFs and those predicted by the surrogate model-based Bayesian inference in Fig.~\ref{PDF_case1} demonstrate the accuracy of the developed approach, as it is also evident from the computed HDRs in Table \ref{EI_HDR_table}. Finally, the Markov chain, and the posterior PDF obtained for the TDS experiment EII are reported in Figs.~\ref{Chain_case_II} and \ref{PDF_case2}, respectively, and the posterior HDR values are reported in Table~\ref{EII_HDR_table}. In this case, the PDFs exhibit one single mode as previously anticipated. This corresponds to the shallow trap ($\overline{{\Delta H}}_1=-30$, $\log(\overline{{N}}_1)=-3$), while the trap in the no-flux regime goes unnoticed. From a Bayesian perspective, this represents an observability limitation of the experiment, being the model of one single trap more likely to represent the material given the experimental evidence. Furthermore, it is noted that the presence of measurement noise does not substantially alter the inference outcome. The modes of the posteriors for the trap densities parameters $\overline{\Delta H}_1$ and $\overline{\Delta H}_2$ of the noise-free scenario are $-29.966$ and $-30.002$, while for the noisy scenario the values $-29.989$ and $-29.961$ are obtained, which represents a difference of $0.077$\% and $0.137$\%, respectively. On the other hand, for parameters $\log(\overline{N}_1)$ and $\log(\overline{N}_2)$ the modes of the posteriors in the noise-free case are $-3.009$ and $-3.016$, whereas in the noisy scenario they take values $-3.017$ and $-3.019$, meaning a difference of $0.266$\% and $0.010$\%, respectively. This confirms that the proposed approach is robust to the presence of measurement noise. Overall, these results illustrate the potential of the developed approach for model selection and information gain analysis of TDS results, which are left for future developments.

\begin{table}
\newcommand\Tstrut{\rule[0.5pt]{0pt}{0.15cm}}         
\newcommand\Bstrut{\rule[-0.15cm]{0pt}{0pt}}   
\footnotesize
\begin{center}
\begin{tabular}{ccccc}
\hline
 HDR & $\overline{\Delta H}_1$ & $\overline{\Delta H}_2$ & log$(\overline{N}_1)$ & log$(\overline{N}_2)$ \Tstrut\Bstrut\\ \hline
80\% HDR (MCMC)       &  \parbox[c][1.1cm]{2.3cm}{(-38.004,-33.084) $\cup$ (-28.931,-22.606)}  & \parbox[c][1.1cm]{2.3cm}{(-37.833,-32.686) $\cup$ (-28.628,-23.134)} & \parbox[c][1.1cm]{2cm}{\centering (-3.075,-2.418)} & \parbox[c][1.1cm]{2cm}{(-3.059,-2.736) $\cup$ (-2.683,-2.409)}\Tstrut \\
80\% HDR (Analytical)       &  \parbox[c][1.1cm]{2.3cm}{ (-38.515,-32.253) $\cup$ (-29.121,-22.729)}  & \parbox[c][1.1cm]{2.3cm}{(-38.179,-32.247) $\cup$ (-29.184,-22.991)} & \parbox[c][1.1cm]{2cm}{\centering (-3.069,-2.386)} & \parbox[c][1.1cm]{2cm}{\centering (-3.099,-2.389)}\Tstrut \\
50\% HDR (MCMC)       &  \parbox[c][1.1cm]{2.3cm}{(-36.726,-33.659) $\cup$ (-27.398,-24.842)}  & \parbox[c][1.1cm]{2.3cm}{(-36.546,-33.280) $\cup$ (-27.143,-25.164)} & \parbox[c][1.1cm]{2cm}{(-2.925,-2.869) $\cup$ (-2.653,-2.427)} & \parbox[c][1.1cm]{2cm}{(-2.898,-2.868) $\cup$ (-2.637,-2.416)}\Tstrut \\
50\% HDR (Analytical)       &  \parbox[c][1.1cm]{2.3cm}{(-36.884,-32.840) $\cup$ (-27.295,-24.816)}  & \parbox[c][1.1cm]{2.3cm}{(-36.680,-32.899) $\cup$ (-27.424,-25.208)} & \parbox[c][1.1cm]{2cm}{\centering (-2.702,-2.406)} & \parbox[c][1.1cm]{2cm}{\centering (-2.703,-2.399)}\Tstrut \\
\hline
\end{tabular}
\caption{HDR at 80\% and 50\% of the PDFs obtained by direct integration and by MCMC for experiment EI.}
\label{EI_HDR_table}
\end{center}
\end{table}


\begin{table}
\newcommand\Tstrut{\rule{0pt}{0.15cm}}         
\newcommand\Bstrut{\rule[-0.15cm]{0pt}{0pt}}   
\small
\begin{center}
\begin{tabular}{ccccc}
\hline
 HDR & $\overline{\Delta H}_1$ & $\overline{\Delta H}_2$ & log$(\overline{N}_1)$ & log$(\overline{N}_2)$ \Tstrut\Bstrut\\
 \hline
80\% HDR (noise-free data)       & (-39.872, -22.145)  & (-34.206,-25.128) & (-5.887,-2.927) & (-5.346,-2.920)\Tstrut \\
80\% HDR (noisy data) & (-39.970,-26.221) & (-39.314,-24.378) & (-5.410,-2.900) & (-5.733,-2.889) \\ 
50\% HDR (noise-free data)       & (-31.526,-28.406)  & (-30.841,-29.199) & (-3.886,-2.924) & (-3.088,-2.946)\Tstrut \\
50\% HDR (noisy data) & (-31.534,-28.698) & (-31.576,-28.656) & (-3.307,-2.896) & (-3.286,-2.883) \\
\hline
\end{tabular}
\caption{HDR at 80\% and 50\% of the PDFs obtained from noisy and noise-free data by MCMC for experiment EII.}
\label{EII_HDR_table}
\end{center}
\end{table}

\section{Conclusions}\label{Sectconc}

This work presents the development of a multi-element PCK meta-model for surrogate model-based Bayesian parameter inference of highly nonlinear engineering models. The proposed  metamodel combines adaptive sparse PCE and Kriging metamodelling to attain both global and local prediction capabilities. The optimal order of the polynomials in the PCE is automatically identified by the LAR algorithm. Then, the optimal PCE is inserted into a Kriging predictor as the trend term, while the stochastic term is fitted through GA optimization. With the aim of tackling non-smoothness in the forward model, a simple regular block partitioning approach has been implemented. On this basis, the space domain is split into a discrete number of subsets where local surrogate models are constructed. Then, the global model response is obtained by combining the local metamodels in a piecewise fashion. Finally, the surrogate model is used for Bayesian parameter estimation using a cost-efficient DRAM MCMC with DR and AM capabilities. The effectiveness of the proposed approach has been validated through two benchmark case studies: (i) the analytical Drop-Wave function; (ii) and a PDE for TDS tests. The presented results and discussion demonstrate the suitability of the proposed scheme to conduct fast Bayesian model estimation of non-linear engineering models. Key findings and contributions of this work include:

\begin{itemize}
	\item Optimal surrogate models ought to be defined by preliminary parametric analyses accounting for prediction accuracy and computational cost. The latter is particularly critical when performing computationally intense applications such as Bayesian parameter estimation. To this aim, this work has presented a set of error metrics and a methodological discussion through two validation case studies.  
    \item The results on the Salomon function have shown that the proposed multi-element PCK model with regular block partitioning provides similar accuracy ($R^2>0.99$, NMAE$<10^{-4}$) as the (classical) PCK approach, while achieving 100 and 10 times shorter evaluation and construction times, respectively. Moreover, the presented results have shown that the proposed method outperforms the SSE technique for the analysis of such a highly nonlinear surface, requiring 20 times fewer samples to achieve a comparable accuracy.
	\item The size of the ED and the number of domain partitions critically determine the computational cost of the developed sparse PCE-Kriging metamodel. Specifically, the partition of non-smooth problems into a finite set of sub-domains allows the sparse adaptive PCE to eliminate a considerable number of high-order components through LAR, so achieving important savings in the construction of the Kriging model and the evaluation of the resulting metamodel. 
	\item The developed surrogate model-based DRAM MCMC approach allows to conduct fast Bayesian parameter inference. In particular, the proposed approach has been applied to the identification of micro-structural traps in metallic alloys subject to TDS. The hydrogen fluxes obtained in TDS test represent a considerable challenge in surrogate modelling due to the presence of diverse regimes depending on the configuration of the hydrogen traps. In terms of $R^2$, the proposed approach is capable of reproducing more than $99.9$\% of the hydrogen diffusion TDS model with computational time savings of 99.3\% with respect to the forward numerical model.
	\item The presented analyses evidence the potential of the developed approach for conducting inverse characterisation of hydrogen-metal interactions. The accuracy of the proposed PCK surrogate model in conjunction with DRAM MCMC opens vast possibilities for future applications in model selection, and information gain analysis of TDS hydrogen desorption tests. 
\end{itemize}

Despite its simplicity, the adopted regular block partitioning model has demonstrated significant performance in terms of computational savings. In this respect, future research will involve the development of more efficient partitioning algorithms that would allow the sampling effort to be localised where the forward model presents greater non-linearities, thus achieving similar accuracies with smaller sample sizes. Another interesting goal for future work consists in the development of multielement surrogate PCK-based models capable of dealing with discontinuities in the response surface.

\section*{Acknowledgements}

This work has been partially supported through the Ministerio de Ciencia e Innovaci\'on [PID2020-116809GB-I00] of Spain and from the Junta de Extremadura through the Research Group Grant [GR18023]. E. Mart\'{\i}nez-Pa\~neda acknowledges financial support from the EPSRC [grant EP/V009680/1] and from UKRI's Future Leaders Fellowship programme [grant MR/V024124/1].

\bibliographystyle{elsarticle-num}
\bibliography{R1_2021_hydrogen_biblio_shortnames}

\end{document}